\documentclass{article}

\usepackage{PRIMEarxiv}

\usepackage{graphicx}
\usepackage[many]{tcolorbox}

\usepackage{mathtools,amssymb,latexsym,amsfonts,stmaryrd,bm}
\usepackage{pbox}
\usepackage{xfrac}
\usepackage{booktabs}
\usepackage{xcolor}
\usepackage{threeparttable}
\usepackage{amsthm}
\usepackage{amsmath}
\usepackage{amssymb}
\usepackage{bbding}
\usepackage{hyperref}
\usepackage{multirow}
\usepackage{tikz}
\usepackage{mathpartir}
\usepackage[ruled,linesnumbered]{algorithm2e}
\usepackage{url}
\usepackage{syntax}
\usepackage{framed}
\usepackage[noend]{algpseudocode}
\usepackage{flushend}
\usepackage{listings}
\usepackage{moresize}
\usepackage{wrapfig}
 
\usepackage{seqsplit}
\usepackage{alltt}
\usepackage{xspace}
\usepackage{enumitem}
\usepackage{pifont}

\usepackage{colortbl}
\usepackage{wrapfig, lipsum}

\usepackage{amsmath, listings, amsthm, amssymb, proof, xspace}

\usepackage{thmtools}

\declaretheoremstyle[spaceabove=\topsep,notefont=\normalfont\itshape]{mystyle}

\newcommand{\revise}[2]{{\color{red}{\ifx&#1&\else- #1\fi}} {\color{ForestGreen}{\ifx&#2&\else+ #2\fi}}}%
\renewcommand{\revise}[2]{#2}%

\usetikzlibrary{arrows,patterns, decorations.pathreplacing}
\newtheorem{definition}{Definition}

\newcommand{\F}{Fig.}
\newcommand{\E}{Eq.}
\newcommand{\T}{Table}
\renewcommand{\S}{Sec.}
\newcommand{\A}{Alg.}

\newcommand{\ignore}[1]{}

\newif\ifshowcomments
\showcommentstrue

\ifshowcomments
\newcommand{\sw}[1]{{\small\color{magenta}{SW:}\emph{\color{magenta}{#1}}}}
\else
\newcommand{\sw}[1]{}
\fi

\ifshowcomments
\newcommand{\jzl}[1]{{\color{pptgreen1}{JZL:}\emph{\color{pptgreen1}{#1}}}}
\else
\newcommand{\jzl}[1]{}
\fi

\lstdefinestyle{base}{
  moredelim=**[is][\color{red}]{@}{@},
  escapeinside={<@}{@>}
}

\newcommand{\parh}[1]{\noindent\textbf{#1}}

\usepackage{amssymb}
\usepackage{pifont}

\newcommand\DejaVuttfamily{%
  \fontfamily{DejaVuSansMono-TLF}\selectfont }

\lstdefinestyle{base}{
  moredelim=**[is][\color{red}]{@}{@},
  escapeinside={<@}{@>}
}

\lstdefinelanguage
   [x64]{Assembler}     
   [x86masm]{Assembler} 
   {morekeywords={CDQE,CQO,CMPSQ,CMPXCHG16B,JRCXZ,LODSQ,MOVSXD, %
                  POPFQ,PUSHFQ,SCASQ,STOSQ,IRETQ,RDTSCP,SWAPGS, %
                  rax,rdx,rcx,rbx,rsi,rdi,rsp,rbp, %
                  r8,r8d,r8w,r8b,r9,r9d,r9w,r9b}} 

\usepackage{listings}
\usepackage{fancyvrb}
\usepackage{color}
\definecolor{lightgray}{rgb}{.9,.9,.9}
\definecolor{darkgray}{rgb}{.4,.4,.4}
\definecolor{purple}{rgb}{0.65, 0.12, 0.82}
\definecolor{commentgreen}{RGB}{63,127,95}

\colorlet{myPurple}{blue!40!red}
\definecolor{myOrange}{RGB}{255,192,0}

\newcommand{\enc}[1]{$\phi^{*}_{\theta}$}
\newcommand{\dec}[1]{$\psi^{*}_{\theta}$}

\lstdefinelanguage{Solidity}{
  keywords={len,delete,int,void,payable, public, event, contract, typeof, new, true, false, catch, function, return, null, catch, switch, var, if, in, while, do, else, case, break,struct,const,socklen_t,sa_familty_t,char,sockaddr},
  keywordstyle=\color{violet}\bfseries,
  ndkeywords={class, export, boolean, throw, implements, import, this},
  ndkeywordstyle=\color{darkgray}\bfseries,
  identifierstyle=\color{black},
  sensitive=false,
  comment=[l]{//},
  escapeinside={(*@}{@*)},          
  morecomment=[s]{/*}{*/},
  commentstyle=\color{commentgreen}\ttfamily,
  stringstyle=\color{red}\ttfamily,
  morestring=[b]',
  morestring=[b]"
}

\newcommand{\rnum}[1]{\uppercase\expandafter{\romannumeral #1\relax}}

\usepackage{xcolor}

\algnewcommand{\LeftComment}[1]{\Statex \(\triangleright\) #1}

\definecolor{pptbrown}{RGB}{132,60,12}
\definecolor{pptgreen}{RGB}{56,87,35}

\let\OLDthebibliography\thebibliography
\renewcommand\thebibliography[1]{
  \OLDthebibliography{#1}
  \setlength{\parskip}{0pt}
  \setlength{\itemsep}{0pt plus 0.1ex}
}

\definecolor{pptgreen}{RGB}{84,130,53}
\definecolor{pptred}{RGB}{176,36,23}
\definecolor{pptblue}{RGB}{68,114,196}

\definecolor{pptgreen1}{RGB}{78,173,91}
\definecolor{pptred1}{RGB}{192,0,0}

\definecolor{pptyellow1}{RGB}{203,195,167}
\definecolor{pptgreen2}{RGB}{184,192,176}
\definecolor{redMark}{RGB}{252, 228, 214}
\definecolor{blueMark}{RGB}{221, 235, 247}
\definecolor{grayMark}{RGB}{242, 242, 242}

\xspace

\SetCommentSty{mycommfont}

\newif\ifshowcomments
\showcommentstrue

\ifshowcomments
\newcommand{\ma}[1]{\mytodoorange{[ma: #1]}}
\else
\newcommand{\ma}[1]{}
\fi

\newcommand{\mytodoorange}[1]{\textcolor{orange}{\ding{46}~{\sf}~#1}}

\usepackage{txfonts}

\title{Benchmarking and Explaining Large Language Model-based Code Generation: A Causality-Centric Approach}

\author{
  Zhenlan Ji, Pingchuan Ma, Zongjie Li, Shuai Wang\thanks{Corresponding author.} \\
  Hong Kong University of Science and Technology \\
  Hong Kong, China\\
  \texttt{\{zjiae, pmaab, zligo, shuaiw\}@cse.ust.hk} \\
}

\begin{document}
\maketitle

\begin{abstract}
  While code generation has been widely used in various software development
  scenarios, the quality of the generated code is not guaranteed. This has been
  a particular concern in the era of large language models (LLMs)-based code
  generation, where LLMs, deemed a complex and powerful black-box model, is
  instructed by a high-level natural language specification, namely a
  \textit{prompt}, to generate code. Nevertheless, effectively evaluating and
  explaining the code generation capability of LLMs is inherently challenging,
  given the complexity of LLMs and the lack of transparency.
  
  Inspired by the recent progress in causality analysis and its application
  in software engineering, this paper launches a \textit{causality
  analysis}-based approach to systematically analyze the causal relations
  between the LLM input prompts and the generated code. To handle various
  technical challenges in this study, we first propose a novel causal
  graph-based representation of the prompt and the generated code, which is
  established over the fine-grained, human-understandable concepts
  in the input prompts. The formed causal graph is then used to identify
  the causal relations between the prompt and the derived code. We
  illustrate the insights that our framework can provide by studying over
  3 popular LLMs with over 12 prompt adjustment
  strategies. 
  The results of these studies illustrate the potential of
  our technique to provide insights into LLM effectiveness, and aid end-users in
  understanding predictions. Additionally, we demonstrate that our approach
  provides actionable insights to improve the quality of the LLM-generated code
  by properly calibrating the prompt. 

\end{abstract}

\section{Introduction}
\label{sec:intro}

Code generation serves as one of the most central problems in automated software
engineering. It enables machines to program automatically to satisfy human
intent expressed in the form of some specification, usually in the form of
natural language. The problem of code generation has been studied for decades,
and has been applied as the basis of many different important domains, such as
program synthesis, program repair, and fuzz testing. In recent years, code
generation has achieved great progress in both academia and industry. In
particular, we observe that large language models (LLMs) have been applied to
support code generation, and have achieved impressive results. For example,
GPT-4 has been applied to generate code from natural language, whose coding
ability is comparable to that of humans~\cite{openai2023gpt}. To date, many LLMs
are already seamlessly integrated into the developer's IDE for commercial usage,
like GitHub Copilot, Amazon CodeWhisperer, and Tabnine.

Despite the great progress, we notice that the quality of the generated code,
especially in the era of LLMs-based code generation, fluctuates under the
variation of natural language specifications. For instance, as will be shown
shortly, the way of expressing the same intent in natural language (i.e.,
prompt) can lead to significantly different code outputs. As a result, the
behavior of LLMs-based code generation is opaque and uninterruptible, which
hinders the broader adoption of LLMs-based code generation in practice. For
instance, it has been reported that GitHub Copilot generates code is highly
unstable~\cite{li2022cctest} or even generates code containing security
vulnerabilities~\cite{pearce2022asleep}.


To date, effectively evaluating and explaining the code generation capability of
LLMs is challenging. While the research community has proposed several
benchmarks to evaluate the code generation capability of LLMs, such as
CodeSearchNet~\cite{husain2019codesearchnet} and
HumanEval~\cite{chen2021evaluating}, those benchmarks focus primarily on the
surface level metrics (e.g., the BLEU score) or functional metrics like 
the pass rate. However, they fail to capture the interplay between the prompt
and the generated code, which is critical to understanding the behavior of
LLMs-based code generation. While some recent work has proposed approaches to
explain the outcome of code generation models~\cite{liu2023reliability}, they
extensively rely on the model's internal mechanics, which are not available for
the most common usage of LLMs-based code generation (i.e., via API). Indeed, to
the best of our knowledge, there is no existing work that systematically
analyzes prompts' impact on the code generated by LLM.

Inspired by the recent progress in causality analysis~\cite{pearl2009causality}
and its application in software
engineering~\cite{sun2022causality,ji2023cc,siebert2023applications}, 
this paper explores a \textit{causality analysis}-based approach to
explaining LLM-based code generation. Our method offers a systematic and
human-understandable explanation, by establishing the causal relations between
the LLM prompts and the generated code. 
Technically, given the text nature of the prompt and code, it is challenging to
represent the prompt and the generated code in a canonical form that is amenable
to causal analysis. To this end, we first propose a novel quantification scheme to
represent the prompt and the generated code via a number of linguistic features
(for prompt) and code features (for generated code). Then, we recast the task of
conducting causality analysis between natural language and code as analyzing
causality analysis among these numerical features. To systematically explore how
the diverse form of prompts affects the generated code, we employ an LLM-based
rephrasing technique (e.g., ``make it longer'', ``make it more technical'') to
generate diverse prompts for the same intent. We then apply a state-of-the-art
causality analysis algorithm, namely DiBS~\cite{lorch2021dibs}, to identify the
causal relations among these features and estimate the average treatment effect
(ATE) of each rephrasing instruction on the generated code. The ATE indicates
the average difference of a feature (e.g., BLEU score) in the generated code
when the prompt is rephrased (e.g., to be lengthy) compared to the original
prompt. We then use the ATE as a building block to enable holistic analysis that
can explain the potential causes on a metric of interest (e.g., BLEU score) in
the generated code. For instance, we can identify which linguistic features in
the prompt that are most likely to cause the LLM to generate high-quality code
and those that are most likely to cause the LLM to generate low-quality code.
This actionable insights provide firm and handy guidance to developers to tune
the prompt to generate high-quality code.

To gauge the effectiveness of our approach, we conduct extensive
experiments using three representative LLMs,
GPT-Neo~\cite{hendrycksapps2021}, GPT-3.5-Turbo (ChatGPT)~\cite{chatgpt},
and GPT-4.
This empirical study enables a comprehensive analysis of the trade-offs and
leads to a number of intriguing findings over LLM prompts and code outputs.
First, as expected, the usage of certain keywords/concepts may notablely
affect the pattern of produced code output, highlighting the need for a
systematic and automated approach to deciding optimal concept/keyword usage
for particular scenarios. 
Second, certain keywords, such as \texttt{Fluent} and \texttt{Formal},
deliver visible ``tradeoffs'' in the generated code, which may become a
practical obstacle for developers to tune the prompt. 
%
Moreover, the reaction of LLMs to prompts with certain properties may be
unreasonable, revealing some severe problems in the LLMs (e.g., overfitting
to the training data). This observation highlights the importance of
taking both human-understandable and other subtle factors into account when
calibrating the LLM prompt. Last, we demonstrate a versatile and important
application of our framework to improve prompts. With empirical assessment,
we show the selected prompts offer superior code generation quality
compared to the original prompt, illustrating the potential of our approach
in assisting developers in refining the prompt.
%
To conclude, this paper makes the following contributions:
\begin{itemize}[leftmargin=*,noitemsep,topsep=0pt]
    \item Given LLM-based code generation, a cornerstone and vastly-used task in
    software engineering, this paper for the first time uses causality analysis
    as a principled approach to analyzing how prompts affect the LLM-based code
    generation process. 
    \item Technically, we propose a set of domain-specific design
    considerations to enable accurate and comprehensive causality analysis in
    the context of LLM-based code generation. We also design a novel approach to
    identify optimal prompts that facilitate instructing LLMs to generate
    high-quality code.
    \item We conduct extensive evaluations over datasets and different LLMs, and
    we obtain actionable suggestions for users to understand and enhance the
    quality of the LLM-based generated code.
\end{itemize}

\parh{Code and Data.}~We provide the code, data, and other supplementary
materials of this research at~\cite{artifact}. We will maintain the released
artifacts to ensure reproducibility and facilitate future research.

\section{Preliminary and Motivation}
\label{sec:background}


\subsection{LLM and Prompt Engineering}
\label{subsec:llm}

\parh{LLM.}~In recent years, LLMs have experienced a surge in popularity
and adoption across various scenarios owing to their promising performance
and high flexiblity on a diverse range of tasks. LLMs are typically trained
on large corpora of text data using self-supervised learning, and can be
fine-tuned on specific downstream tasks with only a few examples (see
below).
To date, LLMs like ChatGPT~\cite{chatgpt}, which contain over 100 billion
parameters, have demonstrated strong capabilities on tasks such as language
translation~\cite{jiao2023chatgpt}, neuron
explanation~\cite{bills2023language}, and even clinical
decision~\cite{singhal2023large}.

\parh{Prompt Engineering.}~One of the key factors that contribute to the success
of LLMs is the input prompt, which is a text or template that provides
task-specific knowledge to the model. The prompt is typically concatenated with
the input text to form the final input to the model. In the task of code
generation, the prompt usually refers to a textual or templated representation
that encompasses the high-level specifications of the desired code.
Conventional ML models are mainly employed following the ``pre-train \&
fine-tune'' paradigm which entails training models on general tasks and
subsequently fine-tuning them for specific downstream tasks.
In contrast, scaled LLMs exhibit properties amenable to few-shot
learning, where prompts can strongly steer model towards producing answers for
desired tasks. This facilitates an era of ``pre-train \& prompt,'' with
properly designed prompts becoming the key to elicit the full potential of LLMs.
To date, various methods that improve LLM performance via prompts have been
proposed, including few-shot
learning~\cite{brown2020language,sanh2021multitask},
chain-of-thought~\cite{wang2022self,wei2022chain},
tree-of-thought~\cite{yao2023tree}, and debate LLM~\cite{du2023improving}.
Consequently, the design and selection of prompts have rapidly gained
importance, spearheading a new research area dubbed ``prompt engineering'' in
LLM research~\cite{liu2021pre}.


\subsection{Code Generation} 
\label{subsec:codegen}

Code generation, also known as program synthesis, is a fundamental software
engineering technique~\cite{gulwani2017program}. Early code generation task was
typically formulated as a search-based problem~\cite{green1981application,
manna1971toward, solar2008program}, with the search space constructed by
constraints. These constraints are derived from the high-level specification and
the target language grammar. Although these approaches can handle simple tasks,
their low generalizability prevents them from being applicable in real-world
circumstances. Deep learning has led to novel neural network-based code
generation methods. The use of various neural architectures such as recurrent
neural networks (RNNs)~\cite{yin2017syntactic}, convolutional neural networks
(CNNs)~\cite{sun2019grammar}, and transformers~\cite{hendrycksapps2021} has
achieved promising results in code generation tasks, including text-to-code and
code-to-code~\cite{lu2021codexglue}. This paper focuses on generating code from
natural language text~\cite{hendrycksapps2021, li2022competition}, with Python3
as the target programming language for its high popularity~\cite{wei2019code,
le2022coderl, fan2023automated}.Our approach, however, is generalizable to
other settings as shown in \S~\ref{sec:discussion}.

Recent years have witnessed a rising interest in applying LLMs to code
generation~\cite{wei2022emergent,li2022cctest,austin2021program}. Although code
generation with LLMs also benefits from prompt engineering, the output quality
is prone to being compromised by the unstable nature of prompts (see example in
\F~\ref{fig:motivation}). Moreover, unlike humans who are capable of correctly
comprehending statements with subtle errors, the low tolerance of compilers and
operating systems for bugs and performance issues further exacerbates the
problem of prompt instability. In addition, the complexity of the natural
language text that constitutes the prompt makes it inherently difficult to
understand which characteristics (e.g., the average length of sentences in the
prompt) of the prompt contribute to the output quality. All of these issues
motivate us to investigate the relations between the prompt and the output
quality of LLMs for code generation.


\subsection{Causality Analysis}
\label{subsec:causality}

As a canonical technique that is extensively applied to analyze complex software
systems~\cite{sun2022causality,zhang2022adaptive,dubslaff2022causality},
causality analysis is proficient at disentangling the complex relations between
various factors into an intuitive causal graph with high interpretability.
Typically, causality analysis comprises two phases: causal discovery and
analysis based on the formed causal graph. 

\parh{Causal Discovery.}~Causal discovery seeks to infer causal relations
between variables from observational data. These inferred causal relations are
often represented as a directed acyclic graph (DAG), referred to as Causal
Graph. Formally,
\begin{definition}[Causal Graph]
    A causal graph is a DAG with nodes $V$ and edges $E$, i.e., $G = (V, E)$,
    where each node $X$ ($X \in V$) represents a random variable and each edge
    $X \rightarrow Y$ ($X, Y \in V$) represents a directed causal relation from
    $X$ to $Y$. $Pa_G(X)$ denotes the parent nodes of $X$ in $G$, i.e., $Pa_G(X)
    = \{N | N \rightarrow X \in E\}$. 
\end{definition}

\begin{definition}[Endogenous and Exogenous Nodes]
    Nodes in the causal graph can be categorized into two distinct groups
    based on the presence or absence of their parent nodes: endogenous
    nodes, whose values are determined by other nodes (i.e., parent nodes)
    in the graph, and exogenous nodes, which derive their values from
    external factors.
\end{definition}

The edges of the causal graph represent \textit{causal relations}, which are
distinct from the commonly known \textit{correlations}. The latter indicates
merely that two variables are statistically correlated, whereas the former
indicates that one variable causes the other. Suppose, for example, that there
are three variables, $X$, $Y$, and $Z$, and that the corresponding causal graph
is $X \leftarrow Z \rightarrow Y$. Here, $Z$ is the common cause of $X$ and $Y$,
also known as a \textit{confounder}. In this case, $X$ and $Y$ are correlated
because they are simultaneously affected by $Z$. $X$ and $Y$ are not causally
related, however, because $X$ is not the cause of $Y$ and vice versa. This
disparity between causal relations and correlation necessitates the use of
causal discovery algorithms. These algorithms can precisely infer the causal
relations between variables from observational data and then construct the
causal graph. 


\parh{Analysis based on Causal Graphs.}~After obtaining the causal graph, further
analysis can be conducted. The prerequisite for the analysis, however, is that
all conclusions derived from the graph's properties must be consistent with the
real world. global Markov assumption~\cite{pearl2009causality,
neal2015introduction}, which is widely assumed in the field of causal
analysis~\cite{spirtes2000causation,pearl2009causality}, resolves this issue.

\begin{definition}[Global Markov Assumption]
    Given a distribution $P$ and its corresponding causal graph $G$, $P$
    and $G$ satisfy the global Markov assumption if $X \Perp_G Y \mid Z
    \implies X \Perp_P Y \mid Z$. 
\end{definition}

$X$, $Y$, and $Z$ are three disjoint sets of variables, which also represent
three nodes/node sets in the causal graph. $X \Perp_P Y \mid Z$ signifies that
$X$ and $Y$ are conditionally independent given $Z$ in $P$, reflecting the
real-world causal relations. And $X \Perp_G Y \mid Z$ denotes that $X$ and $Y$
are separated by $Z$ in $G$, also known as
\textit{d-separation}~\cite{pearl2009causality}. 

On causal graphs, both qualitative and quantitative analysis can be conducted.
Qualitative analysis typically leverages structural information contained within
the causal graph. By traversing the graph, all factors that have a causal effect
on the target variable can be identified. Furthermore, quantitative analysis,
also known as \textit{causal inference}, aims to measure the causal effect of
one variable $X$ on another variable $Y$, by answering \textit{counterfactual}
questions such as ``what would happen to $Y$ if $X$ were set to a different
value?''. In this paper, both qualitative and quantitative analysis are
performed on the causal graph for different purposes, which will be detailed in
\S~\ref{sec:design}.

\section{Research Motivation and Pilot Study}
\label{sec:motivation}

\begin{figure}[!ht]
  \centering
  \includegraphics[width=0.98\linewidth]{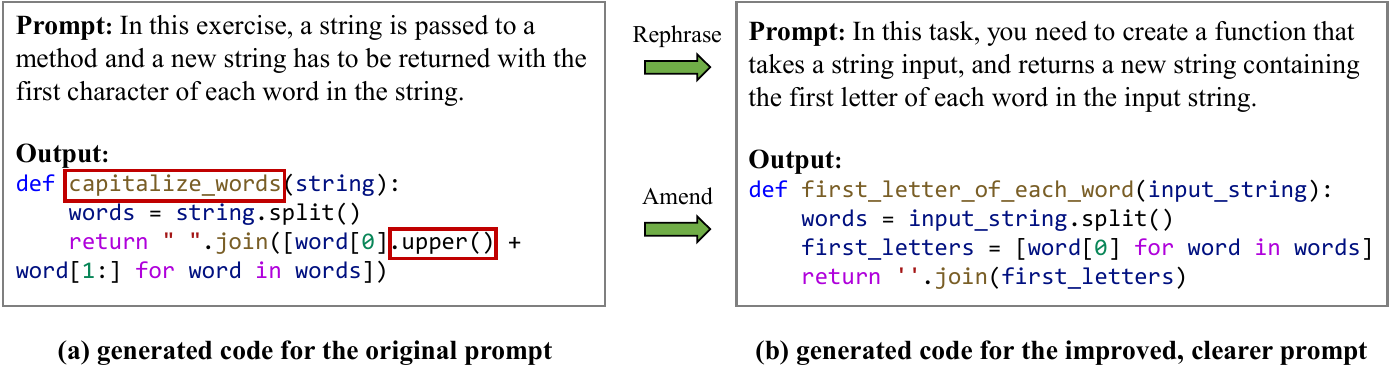}
  \caption{Motivating example of prompt engineering for code generation.
    \textcolor{pptred}{Red boxes} indicate the error in the generated code.}
  \label{fig:motivation}
  \vspace{-10pt}
\end{figure}

\subsection{Effect of Prompt in Code Generation}
\label{subsect:prompt-effect}


As reviewed in \S~\ref{sec:background}, while recent work has shown promising
results in prompt engineering, the design of prompts is still largely a manual
process that relies on human expertise. Thus, evaluating the quality of prompts
remains an open problem~\cite{wang2022no, aggarwal2022towards}, let alone
providing a systematic guideline for prompt design. Overall, considering the
high flexibility of prompts in natural language, we see that the effects of
different types of prompts on LLM performance are not well understood.
Specifically, the criteria for assessing prompt quality are unclear, making it
challenging to determine whether one prompt is superior to another before
actually executing it. Moreover, the effects of prompts are often non-linear and
non-monotonic~\cite{ishibashi2023evaluating}, making it difficult to predict the
performance of a prompt solely based on its design.

\F~\ref{fig:motivation} presents an example of how the performance of a prompt
can be affected by subtle modifications in its design, leading to highly
confusing and non-monotonic results.
In this case, LLM is instructed to generate a program that returns the
first letter of each word in a given string. The relatively unintelligible
prompt in \F~\ref{fig:motivation}(a) misleads the LLM into generating a
program that capitalizes the initial letter of each word. In contrast, the
more explicit instructions provided by the prompt in
\F~\ref{fig:motivation}(b) direct the LLM to generate the correct program.
Evidently, this distinction was not anticipated by
the designer, as the generated code for two semantically equivalent prompts
should be identical. This example illustrates the importance of comprehending
the mechanism of prompts' effect on code generation. Consequently, this work
aims to answer this research question:

\begin{tcolorbox}[size=small]
  How to systematically establish the prompt's influence on LLM-based code
  generation?
\end{tcolorbox}




\subsection{Prompt Adjustment via LLM-Based Rephrasing}
\label{subsect:prompt-adjust}

Following the observation in \F~\ref{fig:motivation}, we interpret that ``prompt
adjustment'' is critical, as a minor change in the prompt may substantially
improve the performance of a given LLM from generating incorrect code
(\F~\ref{fig:motivation}(a)) to correct code (\F~\ref{fig:motivation}(b)).
Thus, the intuition is that by properly adjusting the prompts, we expect to
observe and characterize the relations between the prompt and the generated code
in depth. Nonetheless, prompt adjustment is itself
challenging~\cite{zhou2022large, pryzant2023automatic, yang2023large}.
Generally, a desirable prompt adjustment method shall notably alter one or more
properties of the prompt while preserving its semantics. In natural language
processing, a straightforward method to achieve this goal is to replace words
with their synonyms. However, we argue that this is not an optimal way in our
research context.
Indeed, the frequent occurrence of lengthy text in real-world prompts and
the abundance of synonyms in natural language jointly construct an enormous
search space, making the exhaustive search for the expected prompts
impractical.



\begin{wraptable}{l}{6.7cm}
  \centering
  \caption{Pilot study on the effectiveness of rephrasing in prompt adjustment.}
  \vspace{5pt}
  \resizebox{1.0\linewidth}{!}{
    \begin{tabular}{l|l|l|l}
      \hline
      \textbf{Instruction} & \textbf{Pass Rate} & \textbf{Syntax Error Rate} & \textbf{Stability} \\\hline
      Short                & -0.0029            & +0.0090                    & +0.0534            \\\hline
      Long                 & +0.0095            & -0.0013                    & -0.0184            \\\hline
      Formal               & +0.0061            & -0.0067                    & +0.0011            \\\hline\hline
      \rowcolor{gray!40}
      N/A                  & 0.0000             & +0.0026                    & +0.0290            \\\hline
    \end{tabular}
  }
  \label{tab:pilot-study}
\end{wraptable}

In this work, we propose adjusting the prompt through \textit{rephrasing}. By
explicitly asking an LLM to rephrase the prompt, often depicting a programming
problem, in different manners, we can systematically and smoothly modify the
style of the prompt while keeping its semantics unchanged.
Moreover, the results of a small-scale pilot study, as shown in
\T~\ref{tab:pilot-study}, indicate that rephrasing can effectively affect
the generated code. In this pilot study, we ask ChatGPT~\cite{chatgpt} to
rephrase the given programming problem by following three simple
instructions: make the prompt shorter, longer, or more formal. As a
baseline, we also set up a control group that receives no instructions.
Then, all rephrased programming problems are fed into LLM to generate code,
which is then compared to the code generated from the original prompt. The
results show that rephrasing instructions can indeed affect the generated
code and, more importantly, convey different intentions to the LLM, which
are reflected in the varied change trends of code
metrics.\footnote{In short, the pass rate is the percentage of
  generated code that passes the test cases, the syntax error rate is the
  percentage of generated code that contains syntax errors, and the stability
  is the percentage of generated code that is identical to other code
  generated from the same prompt. To clarify, we aim to demonstrate that
  rephrasing instructions can affect the generated code, as reflected in the
  change trends of these metrics. ``Rephrasing'' however does \textbf{not}
  necessarily mean ``improving'' the generated code.} Thus, this work adopts
LLM-based rephrasing as the basic operator and aims to answer the
following research challenge:

\begin{tcolorbox}[size=small]
  How to effectively explore the prompt space and systematically adjust the
  prompt to achieve the desired effect on the generated code?
\end{tcolorbox}


\subsection{Analysis for Complex Relationships in Code Generation}
\label{subsect:analysis-complex}

\begin{wrapfigure}{l}{0.38\linewidth}
  \centering
  \includegraphics[width=1.0\linewidth]{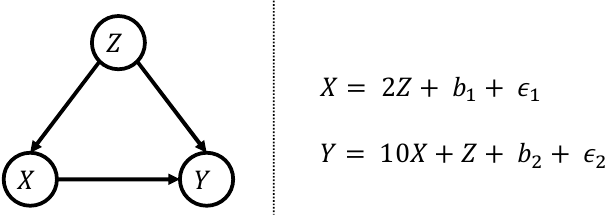}
  \caption{Illustration of the difference between correlation and causation.}
  \label{fig:causation}
\end{wrapfigure}

A straightforward approach to analyzing the relationship between multiple
variables is to identify and calculate the correlation between them. We
argue, however, that correlation is insufficient for analyzing complex
relationships.
Consider an extremely simplified example in \F~\ref{fig:causation}, which
consists of only three variables. Here, $X$ represents the prompt's quality
and $Y$ represents the generated code's quality. Besides, variable $Z$
represents one specific property of the programming question, such as its
difficulty level. Suppose, for instance, these three variables follow the
quantitative relations presented on the right side of
\F~\ref{fig:causation}. $X$'s value is determined by $Z$, and $Y$'s value
is determined by both $X$ and $Z$. In this case, $Z$ is a confounder of the
relation between $X$ and $Y$ because it affects both $X$ and $Y$. Besides,
$b_1$ and $b_2$ are two constants, and $\epsilon_1$ and $\epsilon_2$ denote
random noise with zero mean. The standard procedure for predicting the
value of $Y$ is to train a regression model taking $X$ and $Z$ as inputs.
This trained model is capable of achieving a high level of accuracy, but it
may not capture the correct quantitative relation between $X$ and $Y$. For
example, $Y=21Z+10b_{1}+b2$ is a ``perfect'' model to offer 100\% accuracy
in estimating.\footnote{All noise terms are omitted here because their mean
  is zero.} However, it is evidently not the appropriate model for examining
the effect of the prompt ($X$) on the generated code ($Y$). Any downstream
analysis based on this model will inevitably lead to an incorrect
conclusion that $X$ has no effect on $Y$. This problem will be further
exacerbated in the real-world scenario of LLM-based code generation, where
hundreds of variables, which reflect the properties of the prompt, the
generated code, and the programming question, are involved.


Additionally, the uncertainty of the LLM's output further complicates the
trial-and-error process of empirical analysis~\cite{kuhn2023semantic,
  xiao2022uncertainty, ouyang2023llm}. In particular, the generated code for
a given prompt is stochastic rather than deterministic. Hence, it is
contended that the assessment of the generated code's quality should be
based on the distribution rather than the outcomes of singular or a few
executions.
To conclude,
we attempt to address the following issue in this paper:

\begin{tcolorbox}[size=small]
  How to disentangle the intricate relations between numerous variables in a
  nondeterministic system --- LLM-based code generation --- and to identify the
  correct quantitative relations between prompts and the generated code?
\end{tcolorbox}

\section{Design}
\label{sec:design}

\begin{figure}[!htb]
  \centering
  \includegraphics[width=0.75\linewidth]{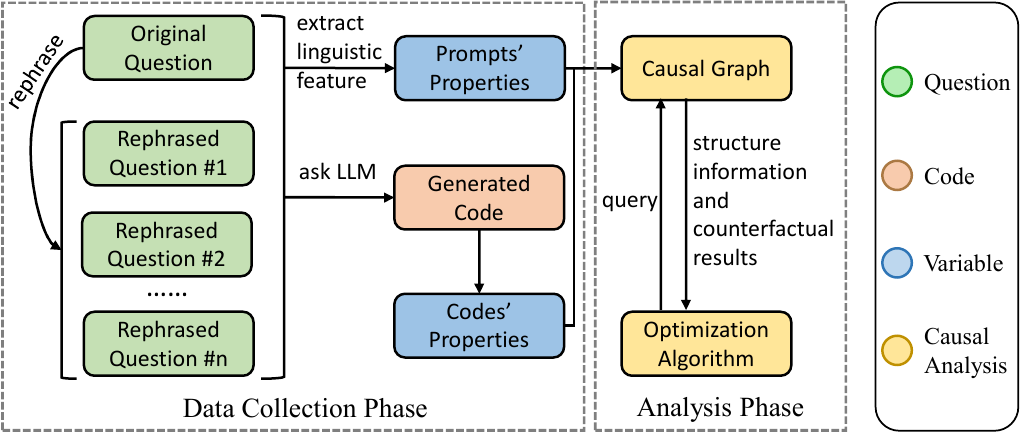}
  \caption{Study overview.}
  \label{fig:workflow}
\end{figure}

\F~\ref{fig:workflow} presents the overview of our study, where two primary
phases are involved: \emph{data collection} and \emph{analysis}. In the
data collection phase, we begin by rephrasing prompts (the green blocks in
\F~\ref{fig:workflow}), which represent questions of programming tasks.
Then, both the rephrased and the original question prompts are fed into two
subtasks. The first subtask is to quantify the characteristics of prompts,
which is accomplished by extracting linguistic features from the natural
language text in these questions. In another subtask, questions are fed
into an LLM to generate code. We further collect a variety of performance
metrics from the generated code, including the code's correctness,
readability, etc. Together, the linguistic features of the prompt and the
performance metrics of the generated code (two blue blocks in
\F~\ref{fig:workflow}) constitute the data for the analysis phase.

During the analysis phase, we first use the causal discovery algorithm to
construct a causal graph that represents the causal relations between all the
variables that we collected. A series of further analyses are conducted on the
causal graph in an effort to identify general principles for code generation
prompt design. Last, a prompt optimization algorithm is employed based on the
causal graph to determine the optimal prompt setting that will generate
high-quality code with the highest probability.

Overall, the workflow in \F~\ref{fig:workflow} is designed to answer the
research questions raised in \S~\ref{subsect:prompt-effect} and address
subsequent technical challenges. Specifically,
\S~\ref{subsect:prompt-quant} proposes a systematic approach for
quantifying the characteristics of prompts. Then,
\S~\ref{subsect:rephrase-gen} addresses the challenge of prompt adjustment
raised in \S~\ref{subsect:prompt-adjust}. Afterwards,
\S~\ref{subsect:causal-analysis} adopt causality analysis to disentangle
the complex relations between the numerous variables involved in this
study. Finally, \S~\ref{subsect:analysis} proposes an algorithm for
learning and explaining prompts' effect on code generation,  thereby
addressing the issue discussed in \S~\ref{subsect:analysis-complex}.


\subsection{Prompt Quantification}
\label{subsect:prompt-quant}

As discussed in \S~\ref{sec:motivation}, the criteria for assessing prompt
quality are generally unclear, making it impractical to compare two prompts
prior to feeding them to LLMs. To address this, we propose to quantify the
characteristics of prompts using linguistic features. Linguistic features are
the building blocks of natural language. They are the elements that make up the
structure of a language, such as morphology, syntax, and semantics. In order to
quantify the prompt for large language models, it is necessary to extract these
linguistic features from the natural language text of the prompts. This can be
done using various techniques, such as part-of-speech
tagging~\cite{barrett2016weakly, xia2019text}, named entity
recognition~\cite{feng2009cognitively}, and dependency
parsing~\cite{manning2014stanford}.

In this paper, we follow Lee et al.~\cite{lee2021pushing, lee2023lftk}, one
state-of-the-art study on linguistic feature in natural language processing, to
extract linguistic features from prompts. Specifically, a total of 255
linguistic features are extracted from prompts, including \textit{lexical}
features (e.g., count of nouns, verbs, adjectives, etc.), \textit{syntactic}
features (e.g., phrasal features~\cite{lu2010automatic}, average height of
parsed dependency tree, etc.), and \textit{semantic} features (e.g., semantic
richness~\cite{lee2021pushing}, word familiarity
features~\cite{collins2014computational}, etc.). The complete list of linguistic
features can be found in the Appendix of Lee et al.~\cite{lee2021pushing}.

By quantifying the linguistic features of a given prompt, we can better
understand the underlying structure and meaning of the text, allowing us to more
accurately capture the nuances and complexities of natural language and
providing a more systematic guideline for prompt design. Further, we presume
that these vast amounts of linguistic features are capable of sufficiently
capturing all interactions between prompts and other variables (including
rephrase instructions and generated code), thereby resolving the unmeasurability
issue of prompt discussed in \S~\ref{sec:motivation}.

\begin{figure}[!htb]
  \centering
  \includegraphics[width=1.0\linewidth]{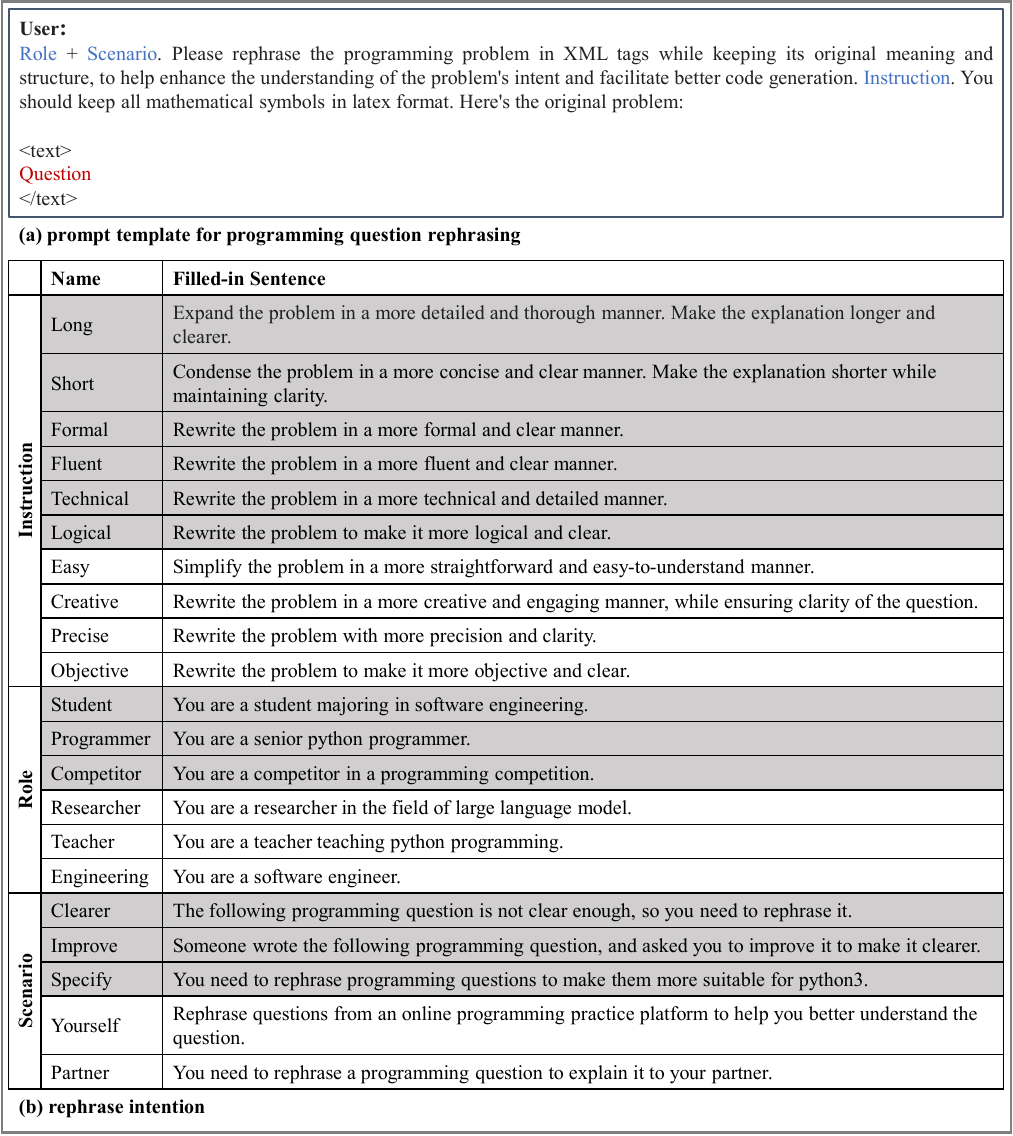}
  \vspace{-10pt}
  \caption{Meta-prompt design. \textcolor{pptred}{Red} text indicates the
    programming question that is filled in by the user, and
    \textcolor{pptblue}{blue} text indicates the pre-defined rephrasing
    intention that is selected by the user.}
  \label{fig:prompt}
  \vspace{-10pt}
\end{figure}

\subsection{Rephrase Generation}
\label{subsect:rephrase-gen}

To explore the impact of prompt design on code generation, we propose rephrasing
the prompt using LLM. In comparison to rule-based mutation, rephrase serves a
more flexible and holistic mutation primitive; it can generate a greater variety
of prompts and substantially improve the prompt diversity. A greater variety of
prompts indicates that the value space for each linguistic feature has been
thoroughly explored. Hence, we presume that the diversity of prompts is
beneficial to the observability of the linguistic features, which is crucial to
the success of subsequent analysis.

Inspired by mature rephrasing tools, such as QuillBot~\cite{quillbot}, we
designed a prompt template for programming question rephrasing, as shown in
\F~\ref{fig:prompt}. Note that, for simplicity, we use \textit{meta-prompt}
to refer to the prompt used for programming question rephrasing. All of
these meta-prompts are refined and optimized by the state-of-the-art LLM,
GPT-4.0, in an effort to enhance the quality of the rephrased prompts and
subsequently escalate the quality of the generated code.

This template also includes a set of rephrasing intentions that can direct
the LLM to rephrase the provided programming question in a particular
manner. These rephrasing intentions are classified as instruction, role,
and scenario. \textit{Instruction} provides the LLM with explicit
instructions, such as \texttt{make it short} and \texttt{make it fluent}.
For \textit{role} and \textit{scenario}, they are designed to provide
context for the LLM, like \texttt{as a student} and \texttt{in a
  programming competition} to unleash the LLM's creativity. Furthermore,
pre-defined intentions from various categories can be combined to form a
more complex rephrasing intention. These rephrasing intentions are intended
to exhaustively cover applicable rephrasing preferences for the programming
question rephrasing task.

We designed a total of 22 rephrasing intentions, which can be combined to form
over one hundred filled-in content for the meta-prompt template; our tentative
study shows that exploring all these combinations results in excessive cost in
the experiment. Therefore, we select six instructions, three roles, and three
scenarios based on the expert's experience and the results of the pilot study.
The selected rephrasing intentions are marked in grey in \F~\ref{fig:prompt}(b).
It is worth noting that the design rephrasing intention is not limited to those
shown in \F~\ref{fig:prompt}(b). Additional rephrasing intention can be added to
the template to expand the diversity of the prompts. This will be discussed in
\S~\ref{sec:discussion}.

\subsection{Causal Analysis}
\label{subsect:causal-analysis}

Causal analysis is a powerful tool for comprehending the intricate relations
between variables. This section describes the methods we employed to construct a
causal graph (causal discovery) and conduct quantitative analysis on the graph
(causal inference).

\begin{figure}[!htb]
  \centering
  \includegraphics[width=0.55\linewidth]{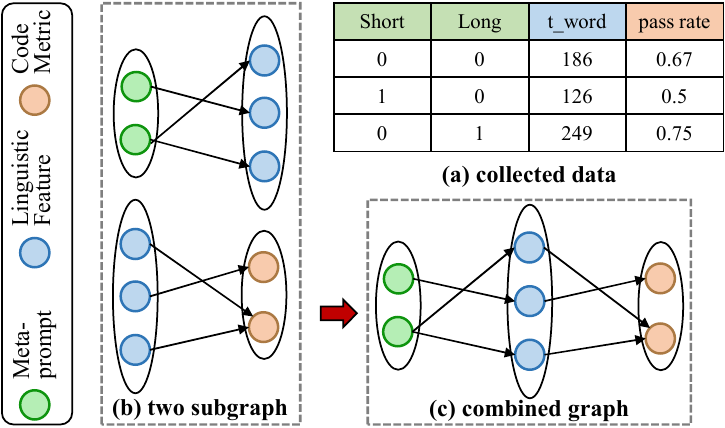}
  \caption{Two-step causal discovery.}
  \label{fig:2step-causal}
  \vspace{-10pt}
\end{figure}

\parh{Causal Discovery.} This study involves a variety of variables, which can
be categorized into three groups: meta-prompt variables (0/1 variables flagging
whether users select a particular rephrasing intention listed in
\F~\ref{fig:prompt}(b) or not), linguistic feature variables, and code metric
variables. By employing causal discovery, it is possible to disentangle the
complex relations between variables and learn a causal graph, where the nodes
represent the variables and the edges represent the causal relations between
nodes. In this study, we employ a state-of-the-art causal discovery algorithm,
DiBS~\cite{lorch2021dibs}. As a gradient-based causal discovery algorithm, DiBS
transforms the causal graph construction procedure into a differentiable
optimization problem, thereby substantially enhancing the efficacy of causal
discovery and ensuring the plausibility of the learned causal
graph~\cite{lorch2021dibs}.

However, the large number of variables involved in this study makes constructing
causal graphs exceedingly difficult. To address this challenge, we propose a
two-step causal discovery approach,
as shown in \F~\ref{fig:2step-causal}.
Holistically, the three aforementioned groups of variables propagate in the
following order: The meta-prompt determines the output of the rephrase
generation task, affecting the linguistic features of the rephrased
programming question, which, in turn, affects the generated code's metrics.
Consequently, the causal graph can be split into two subgraphs, which can
be learned separately. After learning the two subgraphs independently, the
complete causal graph can be obtained by combining them directly.

\parh{Causal Inference.} After learning the causal graph, we conduct
quantitative analysis on the graph to identify the causal effect of one
variable on another. Taking \F~\ref{fig:causation} as an example, the
quantitative analysis seeks to know the magnitude of the change in $Y$ (the
\textit{output} variable) when $X$ (the \textit{treatment} variable)
changes from $x_1$ to $x_2$ while blocking the effect of
\textit{confounding} variable $Z$. Here, $x_1$ and $x_2$ are two specific
values of $X$. One or both of them may not be observed in the data.
Therefore, they are counterfactual values, denoted as $do{X=x_1}$ and
$do{X=x_2}$, respectively. The magnitude of $Y$'s average change reflects
the causal effect of $X$ on $Y$. This is referred to as the \textit{average
  treatment effect} (ATE) within the context of causality
analysis~\cite{pearl2009causality,neal2015introduction}. Below, we present
the formal definition of ATE:

\begin{definition}[ATE]
  ATE of treatment variable $X$ on output variable $Y$ is defined as:
  \begin{equation}
    \textnormal{ATE} = \mathbb{E}[Y \mid do(X=\bm{x}_1)] - \mathbb{E}[Y \mid do(X=\bm{x}_2)]
    \label{eq:ate}
  \end{equation}
  where terms containing $do(\cdot)$ operator are known as \textbf{causal
    estimand}, which cannot be estimated from observational data directly.
\end{definition}

We employ double machine learning (DML)~\cite{chernozhukov2016double} to
estimate causal estimands in \E~\ref{eq:ate}. DML is a state-of-the-art method
and has been extensively applied in the causal analysis
literature~\cite{econml}. It first uses two machine learning models to estimate
treatment variables and output variables, respectively, from confounders. The
causal effect of the treatment variables on the output variables can then be
determined by training another machine learning model that predicts the
relations between the residuals of the two models. In this study, we employ the
DML implementation in the EconML package~\cite{econml}.

\subsection{Establishing Prompt Effect on Code Generation}
\label{subsect:analysis}

\begin{algorithm}[!htbp]
  \scriptsize
  \caption{Analysis}
  \label{alg:analysis}
  \KwIn{Causal Graph $G$, Meta-prompt Variable $M$, Linguistic Features $L_{l} = (L_{1},\cdots,L_{n})$, Code Metrics $C_{l} = (C_{1},\cdots,C_{n})$}
  \KwOut{Dictionary $D$ of Influenced Code Metrics with Explanation for the Effect}
  $D \leftarrow \{\}$\;
  \ForEach{$C \in C_{l}$}{
    $\text{ATE}^{C}_{M} \leftarrow \mathbb{E}[C \mid do(M=1)] - \mathbb{E}[C \mid do(M=0)]$ \tcp*{Computing ATE of $M$ over $C$}
  }
  \tcc{Find the top 3 code metrics that are most affected by $M$}
  Sort $C_{l}$ in descending order according to $\texttt{abs}(\text{ATE}^{C}_{M})$\;
  $C_{S} \leftarrow \texttt{Top3Affected}(C_{l})$\;
  \ForEach{$C \in C_{l}$}{
  $L_{P} \leftarrow \texttt{IdentifyAncestors}(G,C,L_{l})$\;
  \ForEach{$L \in L_{P}$}{
  $l_{M=0} \leftarrow \mathbb{E}[L|M=0]$; $l_{M=1} \leftarrow \mathbb{E}[L|M=1]$\;
  $\text{ATE}^{C}_{L} \leftarrow \mathbb{E}[C \mid do(L=l_{M=1})] - \mathbb{E}[C \mid do(L=l_{M=0})]$\;
  }
  \tcc{Find the linguistic features that are mainly responsible for the effect of $M$ on $C$}
  $D[C] \leftarrow \texttt{SortSelect}(L_{P})$\;
  }

  \Return{$D$}
\end{algorithm}

\A~\ref{alg:analysis} presents the prompt effect analysis procedure. The inputs
for this algorithm are a causal graph $G$, a meta-prompt variable $M$,
linguistic features $L_{l}$, and code metrics $C_{l}$. Typically, the
meta-prompt variable $M$ refers to a binary (0/1) variable flagging if a
particular rephrasing intention is selected or not (``1'' denotes selection).
Moreover, a set of binary variables indicating if multiple rephrasing intentions
are employed or not can also be supported by this algorithm (by invoking
\A~\ref{alg:analysis} for multiple times). Here, the set of binary variables
simulates the situation where a user combines multiple rephrasing intentions to
rephrase a question, as described in \S~\ref{subsect:rephrase-gen}.

Holistically, \A~\ref{alg:analysis} comprises two steps, \textbf{learning}
the effect and \textbf{explaining} the effect. In the first step, we
compute the average treatment effect (ATE) of the meta-prompt variable $M$
on each code metric $C$ (lines 2-4) and then select the top three code
metrics that are most affected by $M$ (lines 5-6). This step is intended to
quantify the extent of $M$'s effect on code generation. In the second
step, we attempt to explain how the changes in meta-prompt $M$ are
propagated to the selected code metrics $C_{S}$ (lines 7-13). Specifically,
we first query the causal graph to find all the ancestor linguistic
features of the chosen code metric $C$ (line 8), as they are potential
mediators of the effect of $M$ on $C$. Then, we intervene on each ancestor
individually to calculate their ATE on $C$ (lines 10-12). Based on the
calculated ATE, we determine the linguistic features primarily responsible
for the effect of $M$ on $C$ (lines 13). Finally, we return a dictionary
$D$ containing the top three influenced code metrics and their
corresponding explanations (line 15).

\section{Experiment Setup}
\label{sec:setup}

Scripts required to conduct the entire study is written in Python3 with about
2.4K lines of code. We run all experiments on a server with AMD Ryzen
3970X CPU and one NVIDIA RTX 3090 GPU.

\subsection{Datasets \& Model}
\label{sec:dataset-model}

\parh{Datasets.}~Our study is conducted on one of the most popular
text-to-code datasets, APPS~\cite{hendrycksapps2021}. It is
exclusively designed for Python code generation and contains 10K Python
problems with difficulty annotations. Each problem is associated with a few
correct solutions and a set of test cases. To learn high-quality causal
graphs, we sample and generate 6K data points from this
dataset.\footnote{Indeed, more data points can be sampled, but we find that
    6K data points are sufficient for our experiments to form causal graphs with reasonable quantity
    and acceptable cost.} Note that
CodeContests~\cite{li2022competition} and other datasets are also
compatible with our framework. The promisingly high generalizability of our framework to
other datasets will be discussed in \S~\ref{sec:discussion}.

\parh{Models.}~Three LLMs are used in our experiments: a pre-trained GPT-Neo
(2.7B) model by Hendrycks et al.~\cite{hendrycksapps2021}, GPT-3.5-Turbo
(ChatGPT), and GPT-4. These three models denote the relatively tiny but powerful
LLM, the most widely-used LLM, and the state-of-the-art LLM, respectively. For
ChatGPT series models (GPT-3.5-Turbo and GPT-4), we use the official API
supported by Azure. Except for the number of generated solutions, which
is set to 3, and the maximum length of generated solutions, which is set to
2,000 to handle long solutions, all parameters are set to their default values.
In this study, each model generates code for the 6K data points and
GPT-3.5-Turbo rephrases the prompt. The reason for this setup is twofold. First,
compared to code generation, prompt rephrasing is a relatively simple task for
for which GPT-3.5-Turbo is sufficient. Our preliminary observation shows that
the more powerful but expensive GPT-4 is an ``overkill'' in this task, while the
tiny GPT-Neo cannot handle it plausibly. Second, in this study, we focus
on the code generation task and the rephrasing itself is quite standard and not
our primary focus.

\subsection{Code Metrics}
\label{sec:code-metric}

Code generation is naturally multi-faceted. For example, generated code should
be correct but fail to adhere to the coding style (e.g., PEP8 for Python). To
comprehensively evaluate the code generation capability of LLMs, we employ a set
of code metrics to measure the quality of generated code. These metrics are
categorized into five categories: correctness, diversity, overhead, readability,
and security, as reported in \T~\ref{tab:code-metric}. The rationale behind this
categorization is that these metrics are orthogonal to each other and can be
used to evaluate the code generation capability of LLMs from different
perspectives. For these metrics, we use the implementation provided
by~\cite{post2018call, wang2021codet5, lu2021codexglue, li2022cctest}.


\begin{table}[!htbp]
    \centering
    \caption{Code metrics employed in this study.}
    \resizebox{0.95\linewidth}{!}{
        \begin{tabular}{l|l|l}
            \hline
            \textbf{Category}                & \textbf{Name} & \textbf{Description}                                                                            \\\hline
            \multirow{5}{*}{Correctness (5)} & pass_rate     & The pass rate of test cases.                                                                    \\\cline{2-3}
                                             & run_err_rate  & The runtime error rate of test cases.                                                           \\\cline{2-3}
                                             & syn_err       & The number of syntax errors revealed by \textit{tree_sitter}~\cite{treesitter}.                 \\\cline{2-3}
                                             & gold_sim_CB   & The similarity (in CodeBLEU~\cite{ren2020codebleu}) between the generated and the ground truth. \\\cline{2-3}
                                             & gold_sim_B    & The similarity (in BLEU~\cite{papineni2002bleu}) between the generated and the ground truth.    \\\hline
            \multirow{2}{*}{Diversity (2)}   & mut_sim_CB    & The mutual similarity (in CodeBLEU) among the generated solutions.                              \\\cline{2-3}
                                             & mut_sim_B     & The mutual similarity (in BLEU) among the generated solutions.                                  \\\hline
            Overhead  (1)                    & timeout_rate  & The timeout rate of test cases.                                                                 \\\hline
            Readability (1)                  & black_count   & The number of places reported by \textit{black}~\cite{Black} where PEP8 is violated.            \\\hline
            Security (1)                     & semgrep_count & The number of potential security bugs revealed by \textit{Semgrep}~\cite{semgrep}.              \\\hline
        \end{tabular}
    }
    \label{tab:code-metric}
\end{table}

\section{Evaluation}
\label{sec:evaluation}

In this section, we evaluate the efficacy and value of our analysis
framework, as described in \S~\ref{sec:design}. To assure the correctness
of subsequent analysis, we first evaluate the accuracy of the learned
causal graphs, which serve as the foundation of our framework
(\textbf{RQ1}). Then, by querying the learned causal graphs, we obtain
several valuable and in-depth insights into LLM-based code generation
(\textbf{RQ2}). Finally, we adopt the learned graphs to guide the
adjustment of the prompt, thereby further demonstrating the value of our
framework (\textbf{RQ3}).

\begin{wraptable}{l}{3.8cm}
    \caption{Statistics of the learned causal graphs.}
    \vspace{5pt}
    \resizebox{1.0\linewidth}{!}{
        \begin{tabular}{l|l|l}
            \hline
                          & nodes & edges \\ \hline
            GPT-Neo       & 45    & 192   \\
            GPT-3.5-Turbo & 50    & 177   \\
            GPT-4         & 45    & 180   \\\hline
        \end{tabular}
    }
    \label{tab:graph}
\end{wraptable}

\T~\ref{tab:graph} reports the statistics of the learned causal graphs. For
causal graph learning, we launch the causal discovery algorithm separately
for each model, since Baluta et al.~\cite{baluta2022membership} have
demonstrated that the causal graphs for distinct models can vary
significantly. For each model, we first collect values for 12 meta-prompt
variables (see \S~\ref{subsect:rephrase-gen}), 255 linguistic variables
(see \S~\ref{subsect:prompt-quant}), and 10 code metrics (see
\S~\ref{tab:code-metric}). The overwhelming number of linguistic variables
is then reduced by removing non-correlated variables, as non-correlation
typically indicates d-separation in the causal graph (refer to
\S~\ref{subsec:causality}). After launching the causal discovery algorithm,
nodes with no edges are also removed for the same reason.

\subsection{RQ1: Verification of the Causal Graphs}
\label{subsec:rq1}

\begin{table}[!htpb]
    \centering
    \scriptsize
    \caption{Predictive capability of the learned causal graphs.}
    \label{tab:rq1-1}
    \resizebox{0.75\linewidth}{!}{
        \begin{tabular}{l||cc|cc|cc}
            \hline
                          & \multicolumn{2}{c}{\textbf{GPT-Neo}} & \multicolumn{2}{c}{\textbf{GPT-3.5-Turbo}} & \multicolumn{2}{c}{\textbf{GPT-4}}                                                \\\cline{2-7}
                          & \textbf{$R^2$}                       & \textbf{MSE}                               & \textbf{$R^2$}                     & \textbf{MSE} & \textbf{$R^2$} & \textbf{MSE} \\\hline\hline
            pass_rate     & 0.8431                               & 0.1527                                     & 0.7600                             & 0.2419       & 0.8206         & 0.1817       \\\hline
            run_err_rate  & 0.8492                               & 0.1510                                     & 0.7978                             & 0.2016       & 0.8474         & 0.1538       \\\hline
            syn_err       & 0.7196                               & 0.2806                                     & 0.7812                             & 0.2156       & 0.8363         & 0.1642       \\\hline
            gold_sim_CB   & 0.9221                               & 0.0783                                     & 0.8432                             & 0.1561       & 0.8708         & 0.1297       \\\hline
            gold_sim_B    & 0.9026                               & 0.0978                                     & 0.8657                             & 0.1358       & 0.8785         & 0.1227       \\\hline
            mut_sim_CB    & 0.9068                               & 0.0923                                     & 0.8654                             & 0.1324       & 0.8878         & 0.1117       \\\hline
            mut_sim_B     & 0.9120                               & 0.0877                                     & 0.8805                             & 0.1194       & 0.8928         & 0.1069       \\\hline
            timeout_rate  & 0.6157                               & 0.3930                                     & 0.7101                             & 0.2916       & 0.6686         & 0.3315       \\\hline
            black_count   & 0.7854                               & 0.2080                                     & 0.8099                             & 0.1913       & 0.8638         & 0.1356       \\\hline
            semgrep_count & 1.0000                               & 0.0000                                     & 0.1812                             & 0.6617       & 0.1448         & 0.9503       \\\hline
        \end{tabular}
    }
\end{table}

As the cornerstone of our analysis framework, the accuracy of the learned
causal graphs is critical. In this section, we endeavor to verify the
accuracy of the learned causal graphs. However, the lack of a ground-truth
causal graph poses an obstacle to verification. To overcome it, we propose a
comprehensive and efficient verification strategy that uses the learned
causal graphs to predict each downstream variable (code metrics listed in
\T~\ref{tab:code-metric} in our case) and then compares the predictions to
the actual values. Idealistically, the learned graph would depict the true
causal relationships between each pair of variables, thereby precisely
depicting the propagation of changes from meta-prompt variables to code
metrics. Consequently, the predictive potential of the learned graph should
correctly reflect its accuracy. For each code metric, we report the $R^2$
score and mean squared error (MSE). The $R^2$ score quantifies the
proportion of the variance in the target variables (code metrics) that can
be predicted from the input variables (meta-prompt variables and the
linguistic properties of the original prompt), with a higher $R^2$ score
indicating a better predictive capability. For MSE, it measures the average
of the squares of the errors; the lower the value, the better.

\T~\ref{tab:rq1-1} reports the verification results. From a holistic
perspective, The high R2 scores and low MSE values indicate that the learned
causal graphs have a strong predictive capability. In most cases, the $R^2$
scores are above 0.8, and the MSE values are below 0.2. The only exception is
the \texttt{semgrep\_count} metric, which represents the number of security
issues detected by Semgrep~\cite{semgrep} in the generated code. On this metric,
the graphs for GPT-3.5-Turbo and GPT-4 perform poorly, whereas the graph for
GPT-Neo attains a perfect score ($R^2$=1.0 and MSE=0.0). We investigate this
phenomenon and discover that the \texttt{semgrep\_count} has a low variance.
Specifically, zero security issues are detected for code generated by GPT-Neo,
while 13 security issues are detected on GPT-3.5-Turbo and four on GPT-4. We
presume that GPT-Neo has been fine-tuned to the point of relative overfitting on
APPS, a clean dataset, resulting in no detected security flaws. For ChatGPT
series models, we attribute the impressive performance to its innovative
training paradigm, reinforcement learning from human feedback
(RLHF)~\cite{ziegler2019fine}. In the subsequent analysis, the
\texttt{semgrep\_count} metric will be omitted as no meaningful insights can be
obtained from it.

\begin{figure}[!ht]
    \centering
    \includegraphics[width=1.0\linewidth]{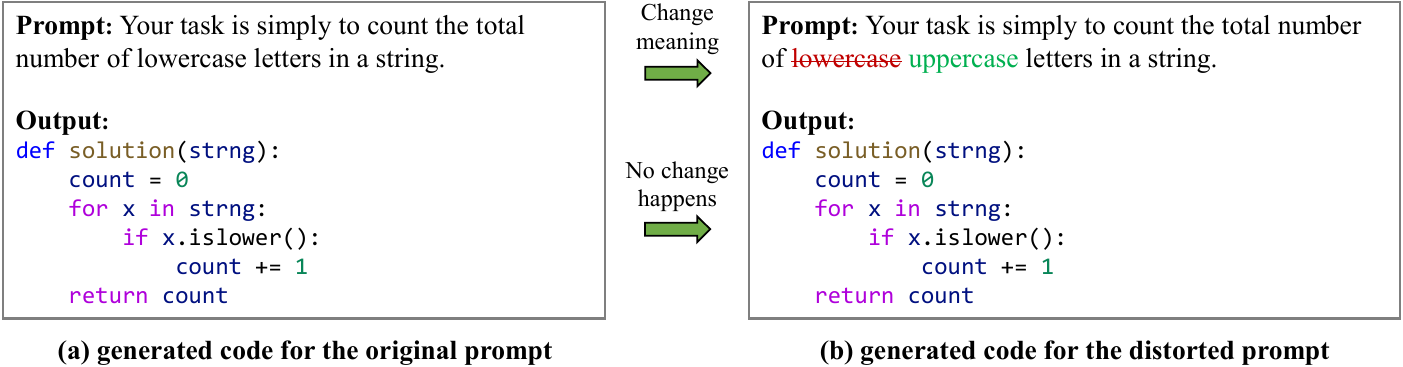}
    \caption{An illustrative example of GPT-Neo's overfitting.}
    \label{fig:overfit}
\end{figure}

There are several other interesting observations when comparing the
performance of the learned graphs for different models. In general, the
graphs for ChatGPT series models have a similar performance pattern, as the
differences between their $R^2$ scores and MSE values are typically small
(less than 0.05). Given that GPT-4.0 is derived from GPT-3.5-Turbo, this is
not surprising. In contrast, the graphs for GPT-Neo display a substantially
distinct performance pattern. In particular, the graph performs
outstandingly on similarity-based metrics (e.g.,
\texttt{gold\_sim\_CodeBLEU}). This phenomenon is interpreted as a
consequence of the overfitting of GPT-Neo on APPS. Instead of ``thinking'' and
then writing code like human programmers, we interpret that GPT-Neo is more likely to
memorize a large number of tiny code snippets from training data and then
combine them according to the requirements. Such characteristics render
GPT-Neo insensitive to changes introduced by meta-prompt variables,
preserving the stability and similarity of the generated code to the ground
truth. \F~\ref{fig:overfit} illustrates this phenomenon. In this example,
altering the count target from \texttt{lowercase} letters to
\texttt{uppercase} letters has no effect on the generated code. This
stickiness to the original ground truth is also reflected in the low
variance of similarity-based metrics, thereby reducing the predictive
difficulty for the learned graph.

\subsection{RQ2: Analysis based on the Causal Graphs}
\label{subsec:rq2}

\begin{figure}[!htbp]
    \centering
    \includegraphics[width=0.9\linewidth]{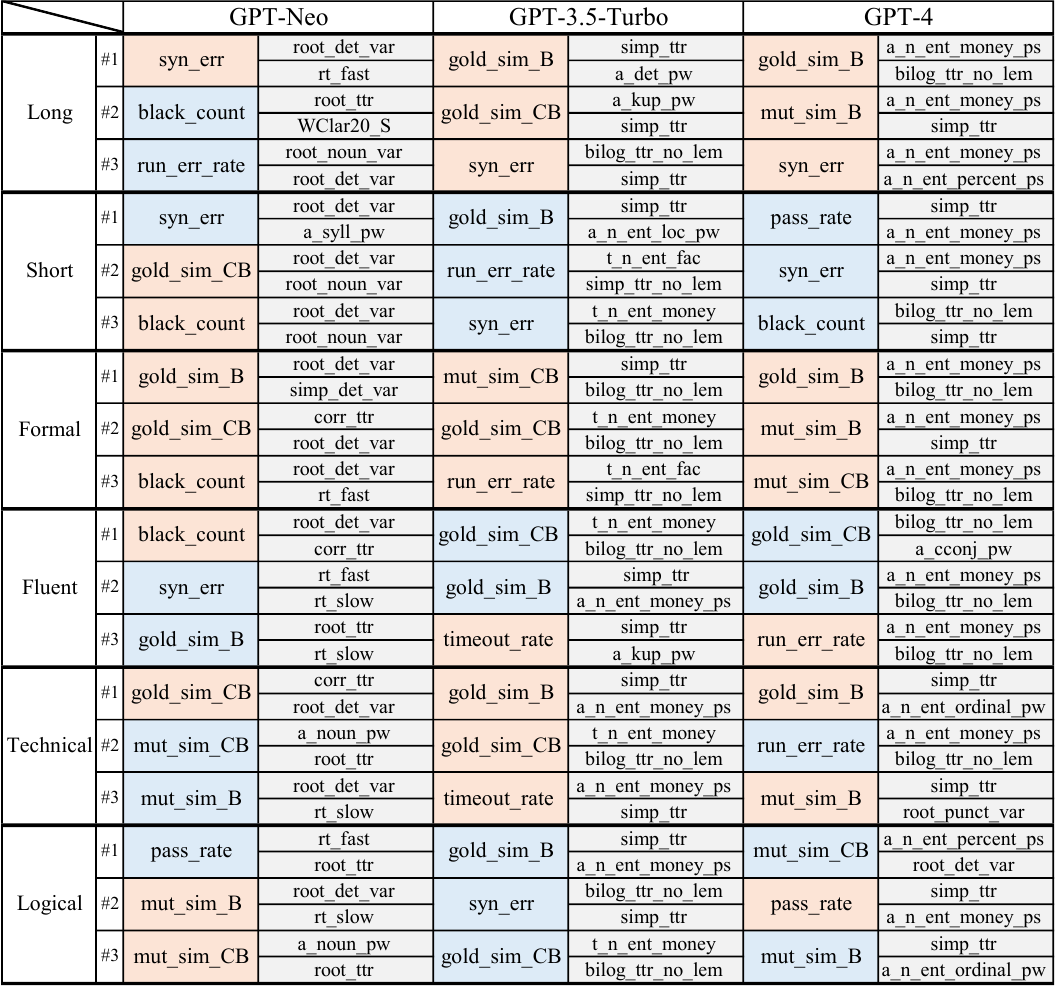}
    \caption{Prompt effect analysis for the \textit{instruction} group. For
        code metrics, \colorbox{redMark}{Red} indicates a negative effect
        received from the meta-prompt variable, while \colorbox{blueMark}{blue}
        indicates a positive effect. The right \colorbox{grayMark}{gray} cells
        record the two primarily responsible linguistic variables for the
        meta-prompt effect.}
    \label{fig:rq2-1}
    \vspace{-10pt}
\end{figure}

\begin{figure}[!htbp]
    \centering
    \includegraphics[width=1.0\linewidth]{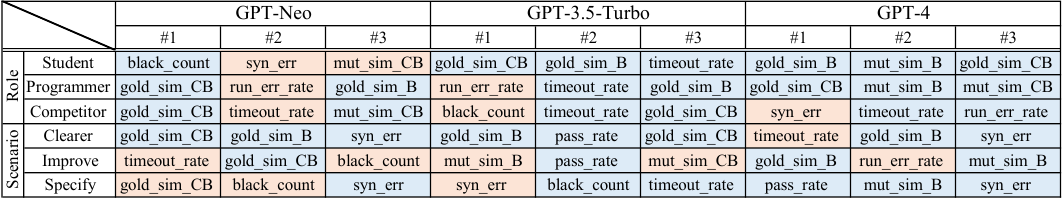}
    \vspace{-10pt}
    \caption{Prompt effect analysis for the \textit{role} and
        \textit{scenario} groups.}
    \label{fig:rq2-2}
\end{figure}

In this section, we assess our analysis framework by employing it to learn
the extent of the impact of meta-prompt variables on code metrics; this
reflects how code metrics can be affected by prompts. \A~\ref{alg:analysis}
is adopted to achieve this objective. Since we divide the meta-prompt
variables into three groups --- \textit{instruction}, \textit{role}, and
\textit{scenario} --- we report the results for each group separately.
Specifically, \F~\ref{fig:rq2-1} presents results for the
\textit{instruction} group, and \F~\ref{fig:rq2-2} shows results for the
\textit{role} and \textit{scenario} groups. Each row in these figures
represents a meta-prompt variable (e.g., \texttt{Long}; see
\F~\ref{fig:prompt} for details), with three sub-rows representing the
three code metrics that are mostly affected by the meta-prompt variable.
Code metrics are highlighted in \colorbox{redMark}{red} and
\colorbox{blueMark}{blue} colors. The former implies that the meta-prompt
variable has a negative effect on the code metric, whereas the latter
indicates a positive effect. Besides, cells marked in
\colorbox{grayMark}{gray} contain the primarily responsible linguistic
variables for their left-hand side code metrics. For simplicity, we attach
the primarily responsible linguistic variables for each code metric only in
\F~\ref{fig:rq2-1}, while omitting them in \F~\ref{fig:rq2-2}. Readers are
encouraged to refer to the repository\cite{artifact} for the complete
results.

The straightforward conclusion from \F~\ref{fig:rq2-1} is that each
meta-prompt variable tends to influence distinct code metrics in different
causal graphs. For example, the mostly affected code metric for meta-prompt
variable \texttt{Short} is \texttt{syn\_err} in GPT-Neo, while it is
\texttt{gold\_sim\_B} in GPT-3.5-Turbo, and \texttt{pass\_rate} in GPT-4.
We attribute this notable difference to the distinction between the three
learned graphs, as mentioned in \T~\ref{tab:graph}'s discussion. However,
there are also some common patterns, particularly in the graphs belonging to
the GPT series models. Typically, the meta-prompt variables \texttt{Long}
and \texttt{Formal} have a negative effect on the generated code,
decreasing its similarity to the ground truth. In contrast, \texttt{Short}
and \texttt{Fluent} are beneficial. In fact, these four meta-prompt variables
constitute two pairs of opposites: \texttt{Long} vs. \texttt{Short} and
\texttt{Formal} vs. \texttt{Fluent}. For \texttt{Long} vs. \texttt{Short},
we attribute this phenomenon to the fact that the former tends to load
rephrasings with unnecessary and even misleading information, whereas the
latter tends to summarize and then extracts the most essential information.
For \texttt{Formal} vs. \texttt{Fluent}, we presume that \texttt{Formal}
complicates the programming logic while \texttt{Fluent} simplifies it.

From the perspective of the linguistic variables, we observe that, in the
majority of instances, certain linguistic variables serve as the primary
responsible variables. \texttt{root\_det\_var} occurs 12 times in the case of
GPT-Neo. This variable represents the value of the total number of unique
determiners divided by the square root of the total number of determiners. The
unexpectedly high frequency of this variable is counterintuitive, as one may not
expect it to have a substantial impact on the generated code. In line with our
discussion in \S~\ref{subsec:rq1}, we assume that this unusual occurrence is
consistent with the overfitting characteristics of GPT-Neo. In contrast,
\texttt{simp\_ttr} and features related to named entities occur most frequently
for GPT series models. The former indicates the degree of lexical variation
of the text, and the latter determines the complexity of the information
contained in the text; both are intuitive and reasonable.

Some additional interesting conclusions can be derived from
\F~\ref{fig:rq2-2}. Compared to the \textit{instruction} group, the effect
of the \textit{role} and \textit{scenario} groups on the generated code in
the cases of GPT series models tends to be more positive. In particular,
these two groups of variables achieve 13 out of 18 positive effects on
GPT-3.5-turbo and 15 out of 18 on GPT-4. This result is consistent with
previous research on prompt engineering~\cite{wei2022chain,du2023improving,
zhang2023cumulative}, indicating that role-playing and scenario setting are
efficient ways to enhance the performance of LLM-based code generation.
GPT-Neo, on the other hand, performs poorly on these two categories of
variables, indicating that role-playing and scenario setting are
inappropriate for models of its size.

\parh{Findings:} There are several interesting findings from the analysis
that are worth mentioning.
\begin{enumerate}[leftmargin=*,noitemsep,topsep=0pt]
    \item Treat the LLM as an ``ignorant idler.'' When you try to improve
          the performance of LLM-based task by rephrasing the prompt, you
          should avoid extending it or stating it in a formal way. Instead,
          you should remove redundant information and make it easier to
          understand.

    \item Take advantage of the analysis pipeline illustrated in this work, and
          watch out for the responsible linguistic variables. When some
          unreasonable linguistic variables frequently work as the primarily
          responsible variables for prompt effect, be careful. This phenomenon
          indicates that the LLM is failing to react reasonably to the prompt
          according to their linguistic properties. In other words, it may be a
          sign of overfitting.

    \item In general, using role-playing and scenario setting denotes a
          universal and effective way to improve the performance of LLM for
          models with sufficient capacity. However, for tiny models, using
          simple rephrasing instructions like ``make it more concise'' can often
          be more effective.
\end{enumerate}


\section{Downstream Application}
\label{sec:application}




In this section, we present a prospective downstream application on the basis of
our approach in \textit{improving the prompt to generate high-quality code}.
Specifically, the procedure is based on the genetic algorithm which is widely
used in the literature of search-based software engineering, such as test case
generation~\cite{varshney2013search}. The genetic algorithm is a metaheuristic
search algorithm that mimics the process of natural selection. It operates on a
population of candidate solutions and iteratively evolves the population to find
the optimal solution. Below, we describe how these procedures are concretized in
our context.

\parh{Fitness Function.}~The goal of the procedure is to find the optimal
rephrasing instructions that maximize the probability of generating high-quality
code. First, as a straightforward solution, we may directly use the objective metric
(e.g., BLEU score) as the fitness function and search for the optimal rephrasing
instructions that maximize the objective metric. However, this solution is
infeasible in our context. The reason is that the objective metric relies on the
ground truth to evaluate the quality of generated code. Even for the metrics
that do not require the ground truth (e.g., \texttt{black\_count}), the solution
is still highly costly as it requires the generation of code for each candidate
solution. In contrast, our causality analysis-based approach actually constitutes a
lightweight surrogate for the objective metric and is free from the ground
truth. Specifically, based on the causal graph, we can estimate the expected
value of the objective metric of a candidate solution without involving the
actual code generation process. This surrogate is much more efficient than the
objective metric and can be used as the fitness function in the genetic
algorithm. More importantly, the surrogate is also expected to be more stable
than actual code generation, as it is asymptotically unbiased and free from the
infamous non-determinism issues in LLM~\cite{ouyang2023llm}.

\parh{Genetic Representation, Modification and Selection.}~We use a genetic
algorithm to find the rephrasing instructions that can maximize a given
objective metric. The instructions are represented as a binary string, with each
bit indicating the selection status of the corresponding instruction. For
example, ``\textit{make it short}'' is represented as \texttt{100000}, and
combining it with ``\textit{make it fluent}'' is \texttt{110000}. The algorithm
operates on these binary strings to find the best instructions. In each
generation, each vector is evaluated using a predefined fitness function. Only
the top-N candidates, based on fitness scores, move to the next generation
(i.e., ``survive''). The new generation is formed using these top-N vectors. Two
standard operators in the genetic algorithm, Crossover and Mutation, are applied
to create new offspring vectors from any two parent vectors. Specifically, two
new vectors are generated using a two-point crossover. In addition to Crossover,
Mutation in our context is implemented to randomly alter bits in a vector to
enhance the population's genetic diversity.

\parh{Result.}~We apply the genetic algorithm to the APPS dataset and use the
BLEU score as the objective metric. Because the rephrasing intentions have been
refined and optimized by GPT-4, we consider single rephrasing instructions to be
an adequate baseline. We evaluate our approach against the best single
rephrasing-intent instruction and the original prompt. The results are shown in
\T~\ref{tab:downstream}. We can see that our approach outperforms the single
rephrasing instruction and the original prompt in most cases. The only exception
is in the case of GPT-Neo, where the overfitting problem may be the cause (see
related discussion in \S~\ref{sec:evaluation}). Overall, we interpret the
results as highly encouraging, suggesting that our algorithm for improving the
prompt to generate high-quality code is effective.

\begin{table}[!htbp]
    \centering
    \vspace{-10pt}
    \caption{Evaluation of the rephrasing prompt generated by our
        algorithm. we highlight the \colorbox{blueMark}{winner}, whose
        BLEU score is the highest and positive. }
    \resizebox{1.0\linewidth}{!}{
        \begin{tabular}{|l||c|c|c|c|c|c|c|c|c|}
            \hline
                     & \multicolumn{3}{c|}{GPT-Neo} & \multicolumn{3}{c|}{GPT-3.5-Turbo} & \multicolumn{3}{c|}{GPT-4}                                                                                                           \\\hline
                     & Original                     & Single                             & Ours                         & Original & Single  & Ours                         & Original & Single  & Ours                         \\\hline
            gold_sim & 0.7071                       & -0.1367                            & -0.0965                      & 0.5014   & +0.0239 & \colorbox{blueMark}{+0.0531} & 0.4867   & +0.0454 & \colorbox{blueMark}{+0.0753} \\\hline
            mut_sim  & 0.8030                       & +0.0338                            & \colorbox{blueMark}{+0.0704} & 0.5072   & +0.0165 & \colorbox{blueMark}{+0.0297} & 0.4971   & +0.0132 & \colorbox{blueMark}{+0.0362} \\\hline
        \end{tabular}
    }
    \label{tab:downstream}
\end{table}

\section{Discussion}
\label{sec:discussion}

\parh{Extensibility.}~While the current technical pipeline is primarily applied
over the task of generating Python3 code, it is extensible to other languages.
Overall, our approach is generally language agnostic; while a few prompts are
language specific, it is easy to see that such prompts can be constructed when
considering other languages. For example, the prompt \texttt{def foo():} can be
replaced with \texttt{function foo() \{} for JavaScript, or \texttt{int foo()
\{} for C. 
Besides, our approach is also extensible to other rephrasing instructions. For
example, we can add the rephrasing instruction \texttt{``make it more
object-oriented''} to the current set of rephrasing instructions. The only
requirement is that the rephrasing instruction should be of reasonable
difficulty for LLMs to understand and implement.

\parh{Threats to Validity.}~Our study is subject to the following threats to
validity. First, our study primarily focuses on the GPT-family models, namely,
GPT-Neo (2.7B), GPT-3.5-Turbo, and GPT-4. While these models are among the most
popular LLMs, there are however other LLMs that are not investigated in this
study such as the T5 model~\cite{raffel2020exploring}. Second, our study is
conducted on APPS~\cite{hendrycksapps2021}, which is currently one of the most
popular datasets for code generation. However, there are other datasets that are
not investigated in this study such as the HumanEval~\cite{chen2021evaluating}
dataset. Third, our study is conducted on Python3 code generation and focuses on
a limited number of rephrasing instructions. However, given the popularity of
Python3 and the extensibility of our approach, we believe our findings are
representative for common usage scenarios and are generalizable to other
languages and rephrasing instructions, as discussed above in ``Extensibility.''

\section{Related Work}
\label{sec:related}


\parh{Theoretical Causality Analysis.}~Causality analysis has been studied in
statistics and machine learning for decades~\cite{pearl2009causality}. In the
past few decades, there has been a surge of interest in learning causal
relations from observational data~\cite{peters2017elements}. Algorithms from
various categories typically employ different strategies for learning the causal
graphs. Constraint-based algorithms are the most classical algorithms in this
area. For instance, the Peter and Clark (PC) algorithm begins with a complete graph and deletes edges using
conditional independence from hypothesis tests~\cite{spirtes2000causation}.
Score-based algorithms recover causal graphs by
maximizing a pre-defined scoring criterion, such as
BDe(u)~\cite{heckerman1995learning}. Representative score-based algorithms
includes HC~\cite{tsamardinos2006max}, GES~\cite{chickering2002learning},
BLIP~\cite{scanagatta2015learning} and GOBNILP~\cite{cussens2017bayesian}.
Recently, gradient-based algorithms have been proposed. For example,
NOTEARS~\cite{zheng2018dags} and DAG-GNN~\cite{yu2019dag} seek to recover the
data generation process while adhering to acyclicity constraints. These
advancements in the theoretical aspect of causality analysis are orthogonal to
our work. In this paper, we focus on the application of it in
evaluating and explaining LLM-based code generation.

\parh{Application of Causality Analysis in Software Engineering.}~Recently, it
has been shown that causality analysis can be applied to various software
engineering tasks~\cite{siebert2023applications}, including
debugging~\cite{fariha2020causality, dubslaff2022causality}, root cause
analysis~\cite{johnson2020causal}, data race
detection~\cite{hsiao2017asyncclock}, and also deep neural network (DNN) testing and
repairing~\cite{sun2022causality, zhang2022adaptive, ji2023cc}. Besides,
causality analysis has been used for understanding empirical software
engineering data and uncover the underlying causal relations. For example,
causality analysis has been used to understand the impact of programming
language on code competitions~\cite{furia2023towards} or various factors that
affect software productivity~\cite{tsunoda2017software}. In this paper, we
advocate the application of causality analysis for understanding the role of
distinct prompts in LLM-based code generation.

\section{Conclusion}

In line with the prosperous adoption of LLMs in code generation, this paper
proposes a novel causality analysis-based approach to systematically analyze the
relations between the LLM input prompts and the generated code. Our solution
facilitates the evaluation and explanation of LLM-based code generation, and
provides actionable insights to improve the quality of the LLM-generated code by
properly calibrating the prompt. Our work can serve as a roadmap for users to 
comprehend and improve the quality of LLM-generated code.

\bibliographystyle{acm}
\bibliography{bib/causality,bib/machine-learning,bib/ref,bib/llm}

\begin{thebibliography}{10}

\bibitem{Black}
Black - the uncompromising code formatter.

\bibitem{semgrep}
Semgrep — find bugs and enforce code standards.

\bibitem{treesitter}
Tree-sitter.

\bibitem{econml}
Econml: A python package for ml-based heterogeneous treatment effects estimation.
\newblock \url{https://github.com/microsoft/EconML}, 2022.

\bibitem{chatgpt}
{ChatGPT}.
\newblock \url{https://chat.openai.com/chat}, 2023.

\bibitem{quillbot}
{Rephrasing Tool - QuillBot AI}.
\newblock \url{https://quillbot.com/}, 2023.

\bibitem{artifact}
Research artifact.
\newblock \url{https://anonymous.4open.science/r/CALL-E7F2/}, 2023.

\bibitem{aggarwal2022towards}
{\sc Aggarwal, A., Sun, J., and Peng, N.}
\newblock Towards robust nlg bias evaluation with syntactically-diverse prompts.
\newblock In {\em Findings of the Association for Computational Linguistics: EMNLP 2022\/} (2022), pp.~6022--6032.

\bibitem{austin2021program}
{\sc Austin, J., Odena, A., Nye, M., Bosma, M., Michalewski, H., Dohan, D., Jiang, E., Cai, C., Terry, M., Le, Q., et~al.}
\newblock Program synthesis with large language models.
\newblock {\em arXiv preprint arXiv:2108.07732\/} (2021).

\bibitem{baluta2022membership}
{\sc Baluta, T., Shen, S., Hitarth, S., Tople, S., and Saxena, P.}
\newblock Membership inference attacks and generalization: A causal perspective.
\newblock {\em arXiv preprint arXiv:2209.08615\/} (2022).

\bibitem{barrett2016weakly}
{\sc Barrett, M., Bingel, J., Keller, F., and S{\o}gaard, A.}
\newblock Weakly supervised part-of-speech tagging using eye-tracking data.
\newblock In {\em Proceedings of the 54th Annual Meeting of the Association for Computational Linguistics (Volume 2: Short Papers)\/} (2016), pp.~579--584.

\bibitem{bills2023language}
{\sc Bills, S., Cammarata, N., Mossing, D., Tillman, H., Gao, L., Goh, G., Sutskever, I., Leike, J., Wu, J., and Saunders, W.}
\newblock Language models can explain neurons in language models.
\newblock {\em \url{https://openaipublic.blob.core.windows.net/neuron-explainer/paper/index.html}\/} (2023).

\bibitem{brown2020language}
{\sc Brown, T., Mann, B., Ryder, N., Subbiah, M., Kaplan, J.~D., Dhariwal, P., Neelakantan, A., Shyam, P., Sastry, G., Askell, A., Agarwal, S., Herbert-Voss, A., Krueger, G., Henighan, T., Child, R., Ramesh, A., Ziegler, D., Wu, J., Winter, C., Hesse, C., Chen, M., Sigler, E., Litwin, M., Gray, S., Chess, B., Clark, J., Berner, C., McCandlish, S., Radford, A., Sutskever, I., and Amodei, D.}
\newblock Language models are few-shot learners.
\newblock In {\em Advances in NeurIPS\/} (2020), H.~Larochelle, M.~Ranzato, R.~Hadsell, M.~Balcan, and H.~Lin, Eds., vol.~33, Curran Associates, Inc.

\bibitem{chen2021evaluating}
{\sc Chen, M., Tworek, J., Jun, H., Yuan, Q., Pinto, H. P. d.~O., Kaplan, J., Edwards, H., Burda, Y., Joseph, N., Brockman, G., et~al.}
\newblock Evaluating large language models trained on code.
\newblock {\em arXiv preprint arXiv:2107.03374\/} (2021).

\bibitem{chernozhukov2016double}
{\sc Chernozhukov, V., Chetverikov, D., Demirer, M., Duflo, E., Hansen, C., Newey, W., and Robins, J.}
\newblock Double/debiased machine learning for treatment and causal parameters.
\newblock {\em arXiv preprint arXiv:1608.00060\/} (2016).

\bibitem{chickering2002learning}
{\sc Chickering, D.~M.}
\newblock Learning equivalence classes of bayesian-network structures.
\newblock {\em Journal of Machine Learning Research 2\/} (2002), 445--498.

\bibitem{collins2014computational}
{\sc Collins-Thompson, K.}
\newblock Computational assessment of text readability: A survey of current and future research.
\newblock {\em ITL-International Journal of Applied Linguistics 165}, 2 (2014), 97--135.

\bibitem{cussens2017bayesian}
{\sc Cussens, J., J{\"a}rvisalo, M., Korhonen, J.~H., and Bartlett, M.}
\newblock Bayesian network structure learning with integer programming: Polytopes, facets and complexity.
\newblock {\em JAIR 58\/} (2017), 185--229.

\bibitem{du2023improving}
{\sc Du, Y., Li, S., Torralba, A., Tenenbaum, J.~B., and Mordatch, I.}
\newblock Improving factuality and reasoning in language models through multiagent debate.
\newblock {\em arXiv preprint arXiv:2305.14325\/} (2023).

\bibitem{dubslaff2022causality}
{\sc Dubslaff, C., Weis, K., Baier, C., and Apel, S.}
\newblock Causality in configurable software systems.
\newblock In {\em Proceedings of the 44th International Conference on Software Engineering\/} (2022), pp.~325--337.

\bibitem{fan2023automated}
{\sc Fan, Z., Gao, X., Mirchev, M., Roychoudhury, A., and Tan, S.~H.}
\newblock Automated repair of programs from large language models.
\newblock In {\em 2023 IEEE/ACM 45th International Conference on Software Engineering (ICSE)\/} (2023), IEEE, pp.~1469--1481.

\bibitem{fariha2020causality}
{\sc Fariha, A., Nath, S., and Meliou, A.}
\newblock Causality-guided adaptive interventional debugging.
\newblock In {\em Proceedings of the 2020 ACM SIGMOD International Conference on Management of Data\/} (2020), pp.~431--446.

\bibitem{feng2009cognitively}
{\sc Feng, L., Elhadad, N., and Huenerfauth, M.}
\newblock Cognitively motivated features for readability assessment.
\newblock In {\em Proceedings of the 12th Conference of the European Chapter of the ACL (EACL 2009)\/} (2009), pp.~229--237.

\bibitem{furia2023towards}
{\sc Furia, C.~A., Torkar, R., and Feldt, R.}
\newblock Towards causal analysis of empirical software engineering data: The impact of programming languages on coding competitions.
\newblock {\em arXiv preprint arXiv:2301.07524\/} (2023).

\bibitem{green1981application}
{\sc Green, C.}
\newblock Application of theorem proving to problem solving.
\newblock In {\em Readings in Artificial Intelligence}. Elsevier, 1981, pp.~202--222.

\bibitem{gulwani2017program}
{\sc Gulwani, S., Polozov, O., Singh, R., et~al.}
\newblock Program synthesis.
\newblock {\em Foundations and Trends{\textregistered} in Programming Languages 4}, 1-2 (2017), 1--119.

\bibitem{heckerman1995learning}
{\sc Heckerman, D., Geiger, D., and Chickering, D.~M.}
\newblock Learning bayesian networks: The combination of knowledge and statistical data.
\newblock {\em Machine Learning\/} (1995).

\bibitem{hendrycksapps2021}
{\sc Hendrycks, D., Basart, S., Kadavath, S., Mazeika, M., Arora, A., Guo, E., Burns, C., Puranik, S., He, H., Song, D., and Steinhardt, J.}
\newblock Measuring coding challenge competence with apps.
\newblock {\em NeurIPS\/} (2021).

\bibitem{hsiao2017asyncclock}
{\sc Hsiao, C.-H., Narayanasamy, S., Khan, E. M.~I., Pereira, C.~L., and Pokam, G.~A.}
\newblock Asyncclock: Scalable inference of asynchronous event causality.
\newblock {\em ACM SIGPLAN Notices 52}, 4 (2017), 193--205.

\bibitem{husain2019codesearchnet}
{\sc Husain, H., Wu, H.-H., Gazit, T., Allamanis, M., and Brockschmidt, M.}
\newblock Codesearchnet challenge: Evaluating the state of semantic code search.
\newblock {\em arXiv preprint arXiv:1909.09436\/} (2019).

\bibitem{ishibashi2023evaluating}
{\sc Ishibashi, Y., Bollegala, D., Sudoh, K., and Nakamura, S.}
\newblock Evaluating the robustness of discrete prompts.
\newblock In {\em Proceedings of the 17th Conference of the European Chapter of the Association for Computational Linguistics\/} (2023), pp.~2365--2376.

\bibitem{ji2023cc}
{\sc Ji, Z., Ma, P., Yuan, Y., and Wang, S.}
\newblock Cc: Causality-aware coverage criterion for deep neural networks.
\newblock In {\em 2023 IEEE/ACM 45th International Conference on Software Engineering (ICSE)\/} (2023), IEEE, pp.~1788--1800.

\bibitem{jiao2023chatgpt}
{\sc Jiao, W., Wang, W., Huang, J.-t., Wang, X., and Tu, Z.}
\newblock Is chatgpt a good translator? a preliminary study.
\newblock {\em arXiv preprint arXiv:2301.08745\/} (2023).

\bibitem{johnson2020causal}
{\sc Johnson, B., Brun, Y., and Meliou, A.}
\newblock Causal testing: understanding defects' root causes.
\newblock In {\em Proceedings of the ACM/IEEE 42nd International Conference on Software Engineering\/} (2020), pp.~87--99.

\bibitem{kuhn2023semantic}
{\sc Kuhn, L., Gal, Y., and Farquhar, S.}
\newblock Semantic uncertainty: Linguistic invariances for uncertainty estimation in natural language generation.
\newblock {\em arXiv preprint arXiv:2302.09664\/} (2023).

\bibitem{le2022coderl}
{\sc Le, H., Wang, Y., Gotmare, A.~D., Savarese, S., and Hoi, S. C.~H.}
\newblock Coderl: Mastering code generation through pretrained models and deep reinforcement learning.
\newblock {\em Advances in Neural Information Processing Systems 35\/} (2022), 21314--21328.

\bibitem{lee2021pushing}
{\sc Lee, B.~W., Jang, Y.~S., and Lee, J.}
\newblock Pushing on text readability assessment: A transformer meets handcrafted linguistic features.
\newblock In {\em Proceedings of the 2021 Conference on Empirical Methods in Natural Language Processing\/} (2021), pp.~10669--10686.

\bibitem{lee2023lftk}
{\sc Lee, B.~W., and Lee, J.}
\newblock {LFTK}: Handcrafted features in computational linguistics.
\newblock In {\em Proceedings of the 18th Workshop on Innovative Use of NLP for Building Educational Applications (BEA 2023)\/} (July 2023), pp.~1--19.

\bibitem{li2022competition}
{\sc Li, Y., Choi, D., Chung, J., Kushman, N., Schrittwieser, J., Leblond, R., Eccles, T., Keeling, J., Gimeno, F., Dal~Lago, A., Hubert, T., Choy, P., de~Masson~d'Autume, C., Babuschkin, I., Chen, X., Huang, P.-S., Welbl, J., Gowal, S., Cherepanov, A., Molloy, J., Mankowitz, D., Sutherland~Robson, E., Kohli, P., de~Freitas, N., Kavukcuoglu, K., and Vinyals, O.}
\newblock Competition-level code generation with alphacode.
\newblock {\em arXiv preprint arXiv:2203.07814\/} (2022).

\bibitem{li2022cctest}
{\sc Li, Z., Wang, C., Liu, Z., Wang, H., Wang, S., and Gao, C.}
\newblock Cctest: Testing and repairing code completion systems, 2023.

\bibitem{liu2021pre}
{\sc Liu, P., Yuan, W., Fu, J., Jiang, Z., Hayashi, H., and Neubig, G.}
\newblock Pre-train, prompt, and predict: {A} systematic survey of prompting methods in natural language processing.
\newblock {\em {ACM} Comput. Surv.\/} (2023).

\bibitem{liu2023reliability}
{\sc Liu, Y., Tantithamthavorn, C., Liu, Y., and Li, L.}
\newblock On the reliability and explainability of automated code generation approaches.
\newblock {\em arXiv preprint arXiv:2302.09587\/} (2023).

\bibitem{lorch2021dibs}
{\sc Lorch, L., Rothfuss, J., Sch{\"o}lkopf, B., and Krause, A.}
\newblock Dibs: Differentiable bayesian structure learning.
\newblock {\em Advances in Neural Information Processing Systems 34\/} (2021), 24111--24123.

\bibitem{lu2021codexglue}
{\sc Lu, S., Guo, D., Ren, S., Huang, J., Svyatkovskiy, A., Blanco, A., Clement, C., Drain, D., Jiang, D., Tang, D., et~al.}
\newblock Codexglue: A machine learning benchmark dataset for code understanding and generation.
\newblock {\em arXiv preprint arXiv:2102.04664\/} (2021).

\bibitem{lu2010automatic}
{\sc Lu, X.}
\newblock Automatic analysis of syntactic complexity in second language writing.
\newblock {\em International journal of corpus linguistics 15}, 4 (2010), 474--496.

\bibitem{manna1971toward}
{\sc Manna, Z., and Waldinger, R.~J.}
\newblock Toward automatic program synthesis.
\newblock {\em Communications of the ACM 14}, 3 (1971), 151--165.

\bibitem{manning2014stanford}
{\sc Manning, C.~D., Surdeanu, M., Bauer, J., Finkel, J.~R., Bethard, S., and McClosky, D.}
\newblock The stanford corenlp natural language processing toolkit.
\newblock In {\em Proceedings of 52nd annual meeting of the association for computational linguistics: system demonstrations\/} (2014), pp.~55--60.

\bibitem{neal2015introduction}
{\sc Neal, B.}
\newblock Introduction to causal inference.

\bibitem{openai2023gpt}
{\sc OpenAI, R.}
\newblock Gpt-4 technical report.
\newblock {\em arXiv\/} (2023), 2303--08774.

\bibitem{ouyang2023llm}
{\sc Ouyang, S., Zhang, J.~M., Harman, M., and Wang, M.}
\newblock Llm is like a box of chocolates: the non-determinism of chatgpt in code generation.
\newblock {\em arXiv preprint arXiv:2308.02828\/} (2023).

\bibitem{papineni2002bleu}
{\sc Papineni, K., Roukos, S., Ward, T., and Zhu, W.-J.}
\newblock Bleu: a method for automatic evaluation of machine translation.
\newblock In {\em Proceedings of the 40th annual meeting of the Association for Computational Linguistics\/} (2002), pp.~311--318.

\bibitem{pearce2022asleep}
{\sc Pearce, H., Ahmad, B., Tan, B., Dolan-Gavitt, B., and Karri, R.}
\newblock Asleep at the keyboard? assessing the security of github copilot’s code contributions.
\newblock In {\em 2022 IEEE Symposium on Security and Privacy (SP)\/} (2022), IEEE, pp.~754--768.

\bibitem{pearl2009causality}
{\sc Pearl, J.}
\newblock {\em Causality}.
\newblock Cambridge university press, 2009.

\bibitem{peters2017elements}
{\sc Peters, J., Janzing, D., and Sch{\"o}lkopf, B.}
\newblock {\em Elements of causal inference: foundations and learning algorithms}.
\newblock The MIT Press, 2017.

\bibitem{post2018call}
{\sc Post, M.}
\newblock A call for clarity in reporting {BLEU} scores.
\newblock In {\em Proceedings of the Third Conference on Machine Translation: Research Papers\/} (Belgium, Brussels, Oct. 2018), Association for Computational Linguistics, pp.~186--191.

\bibitem{pryzant2023automatic}
{\sc Pryzant, R., Iter, D., Li, J., Lee, Y.~T., Zhu, C., and Zeng, M.}
\newblock Automatic prompt optimization with" gradient descent" and beam search.
\newblock {\em arXiv preprint arXiv:2305.03495\/} (2023).

\bibitem{raffel2020exploring}
{\sc Raffel, C., Shazeer, N., Roberts, A., Lee, K., Narang, S., Matena, M., Zhou, Y., Li, W., and Liu, P.~J.}
\newblock Exploring the limits of transfer learning with a unified text-to-text transformer.
\newblock {\em The Journal of Machine Learning Research 21}, 1 (2020), 5485--5551.

\bibitem{ren2020codebleu}
{\sc Ren, S., Guo, D., Lu, S., Zhou, L., Liu, S., Tang, D., Sundaresan, N., Zhou, M., Blanco, A., and Ma, S.}
\newblock Codebleu: a method for automatic evaluation of code synthesis.
\newblock {\em arXiv preprint arXiv:2009.10297\/} (2020).

\bibitem{sanh2021multitask}
{\sc Sanh, V., Webson, A., Raffel, C., Bach, S.~H., Sutawika, L., Alyafeai, Z., Chaffin, A., Stiegler, A., Raja, A., Dey, M., Bari, M.~S., Xu, C., Thakker, U., Sharma, S.~S., Szczechla, E., Kim, T., Chhablani, G., Nayak, N.~V., Datta, D., Chang, J., Jiang, M.~T., Wang, H., Manica, M., Shen, S., Yong, Z.~X., Pandey, H., Bawden, R., Wang, T., Neeraj, T., Rozen, J., Sharma, A., Santilli, A., F{\'{e}}vry, T., Fries, J.~A., Teehan, R., Scao, T.~L., Biderman, S., Gao, L., Wolf, T., and Rush, A.~M.}
\newblock Multitask prompted training enables zero-shot task generalization.
\newblock In {\em {ICLR} 2022, Virtual Event, April 25-29, 2022\/} (2022).

\bibitem{scanagatta2015learning}
{\sc Scanagatta, M., de~Campos, C.~P., Corani, G., and Zaffalon, M.}
\newblock Learning bayesian networks with thousands of variables.
\newblock In {\em NIPS\/} (2015), pp.~1864--1872.

\bibitem{siebert2023applications}
{\sc Siebert, J.}
\newblock Applications of statistical causal inference in software engineering.
\newblock {\em Information and Software Technology\/} (2023), 107198.

\bibitem{singhal2023large}
{\sc Singhal, K., Azizi, S., Tu, T., Mahdavi, S.~S., Wei, J., Chung, H.~W., Scales, N., Tanwani, A., Cole-Lewis, H., Pfohl, S., et~al.}
\newblock Large language models encode clinical knowledge.
\newblock {\em Nature\/} (2023), 1--9.

\bibitem{solar2008program}
{\sc Solar-Lezama, A.}
\newblock {\em Program synthesis by sketching}.
\newblock University of California, Berkeley, 2008.

\bibitem{spirtes2000causation}
{\sc Spirtes, P., Glymour, C.~N., Scheines, R., and Heckerman, D.}
\newblock {\em Causation, prediction, and search}.
\newblock 2000.

\bibitem{sun2022causality}
{\sc Sun, B., Sun, J., Pham, L.~H., and Shi, J.}
\newblock Causality-based neural network repair.
\newblock In {\em Proceedings of the 44th International Conference on Software Engineering\/} (2022), pp.~338--349.

\bibitem{sun2019grammar}
{\sc Sun, Z., Zhu, Q., Mou, L., Xiong, Y., Li, G., and Zhang, L.}
\newblock A grammar-based structural cnn decoder for code generation.
\newblock In {\em Proceedings of the AAAI conference on artificial intelligence\/} (2019), vol.~33, pp.~7055--7062.

\bibitem{tsamardinos2006max}
{\sc Tsamardinos, I., Brown, L.~E., and Aliferis, C.~F.}
\newblock The max-min hill-climbing bayesian network structure learning algorithm.
\newblock {\em Machine learning 65}, 1 (2006), 31--78.

\bibitem{tsunoda2017software}
{\sc Tsunoda, M., and Amasaki, S.}
\newblock On software productivity analysis with propensity score matching.
\newblock In {\em 2017 ACM/IEEE International Symposium on Empirical Software Engineering and Measurement (ESEM)\/} (2017), IEEE, pp.~436--441.

\bibitem{varshney2013search}
{\sc Varshney, S., and Mehrotra, M.}
\newblock Search based software test data generation for structural testing: a perspective.
\newblock {\em ACM SIGSOFT Software Engineering Notes 38}, 4 (2013), 1--6.

\bibitem{wang2022no}
{\sc Wang, C., Yang, Y., Gao, C., Peng, Y., Zhang, H., and Lyu, M.~R.}
\newblock No more fine-tuning? an experimental evaluation of prompt tuning in code intelligence.
\newblock In {\em Proceedings of the 30th ACM Joint European Software Engineering Conference and Symposium on the Foundations of Software Engineering\/} (2022), pp.~382--394.

\bibitem{wang2022self}
{\sc Wang, X., Wei, J., Schuurmans, D., Le, Q.~V., Chi, E.~H., and Zhou, D.}
\newblock Self-consistency improves chain of thought reasoning in language models.
\newblock {\em CoRR abs/2203.11171\/} (2022).

\bibitem{wang2021codet5}
{\sc Wang, Y., Wang, W., Joty, S., and Hoi, S.~C.}
\newblock Codet5: Identifier-aware unified pre-trained encoder-decoder models for code understanding and generation.
\newblock {\em arXiv preprint arXiv:2109.00859\/} (2021).

\bibitem{wei2019code}
{\sc Wei, B., Li, G., Xia, X., Fu, Z., and Jin, Z.}
\newblock Code generation as a dual task of code summarization.
\newblock {\em Advances in neural information processing systems 32\/} (2019).

\bibitem{wei2022emergent}
{\sc Wei, J., Tay, Y., Bommasani, R., Raffel, C., Zoph, B., Borgeaud, S., Yogatama, D., Bosma, M., Zhou, D., Metzler, D., et~al.}
\newblock Emergent abilities of large language models.
\newblock {\em arXiv preprint arXiv:2206.07682\/} (2022).

\bibitem{wei2022chain}
{\sc Wei, J., Wang, X., Schuurmans, D., Bosma, M., Chi, E., Le, Q., and Zhou, D.}
\newblock Chain of thought prompting elicits reasoning in large language models.
\newblock {\em arXiv preprint arXiv:2201.11903\/} (2022).

\bibitem{xia2019text}
{\sc Xia, M., Kochmar, E., and Briscoe, T.}
\newblock Text readability assessment for second language learners.
\newblock {\em arXiv preprint arXiv:1906.07580\/} (2019).

\bibitem{xiao2022uncertainty}
{\sc Xiao, Y., Liang, P.~P., Bhatt, U., Neiswanger, W., Salakhutdinov, R., and Morency, L.-P.}
\newblock Uncertainty quantification with pre-trained language models: A large-scale empirical analysis.
\newblock {\em arXiv preprint arXiv:2210.04714\/} (2022).

\bibitem{yang2023large}
{\sc Yang, C., Wang, X., Lu, Y., Liu, H., Le, Q.~V., Zhou, D., and Chen, X.}
\newblock Large language models as optimizers.
\newblock {\em arXiv preprint arXiv:2309.03409\/} (2023).

\bibitem{yao2023tree}
{\sc Yao, S., Yu, D., Zhao, J., Shafran, I., Griffiths, T.~L., Cao, Y., and Narasimhan, K.}
\newblock Tree of thoughts: Deliberate problem solving with large language models.
\newblock {\em arXiv preprint arXiv:2305.10601\/} (2023).

\bibitem{yin2017syntactic}
{\sc Yin, P., and Neubig, G.}
\newblock A syntactic neural model for general-purpose code generation.
\newblock {\em arXiv preprint arXiv:1704.01696\/} (2017).

\bibitem{yu2019dag}
{\sc Yu, Y., Chen, J., Gao, T., and Yu, M.}
\newblock Dag-gnn: Dag structure learning with graph neural networks.
\newblock In {\em International Conference on Machine Learning\/} (2019), PMLR, pp.~7154--7163.

\bibitem{zhang2022adaptive}
{\sc Zhang, M., and Sun, J.}
\newblock Adaptive fairness improvement based on causality analysis.
\newblock In {\em Proceedings of the 30th ACM Joint European Software Engineering Conference and Symposium on the Foundations of Software Engineering\/} (2022), pp.~6--17.

\bibitem{zhang2023cumulative}
{\sc Zhang, Y., Yang, J., Yuan, Y., and Yao, A. C.-C.}
\newblock Cumulative reasoning with large language models.
\newblock {\em arXiv preprint arXiv:2308.04371\/} (2023).

\bibitem{zheng2018dags}
{\sc Zheng, X., Aragam, B., Ravikumar, P.~K., and Xing, E.~P.}
\newblock Dags with no tears: Continuous optimization for structure learning.
\newblock {\em Advances in Neural Information Processing Systems 31\/} (2018).

\bibitem{zhou2022large}
{\sc Zhou, Y., Muresanu, A.~I., Han, Z., Paster, K., Pitis, S., Chan, H., and Ba, J.}
\newblock Large language models are human-level prompt engineers.
\newblock {\em arXiv preprint arXiv:2211.01910\/} (2022).

\bibitem{ziegler2019fine}
{\sc Ziegler, D.~M., Stiennon, N., Wu, J., Brown, T.~B., Radford, A., Amodei, D., Christiano, P., and Irving, G.}
\newblock Fine-tuning language models from human preferences.
\newblock {\em arXiv preprint arXiv:1909.08593\/} (2019).

\end{thebibliography}

\end{document}